\documentclass[twocolumn,showpacs,preprintnumbers,amsmath,amssymb]{revtex4}

\usepackage{graphicx}
\usepackage{dcolumn}
\usepackage{bm}
\usepackage{graphics}
\usepackage{latexsym}
\usepackage{graphics}
\usepackage{dcolumn}
\usepackage{epsfig}
\usepackage{amssymb} 
\usepackage{amsmath} 
\usepackage{rotating} 
\usepackage{rotate} 
\usepackage{color} 
\usepackage{graphicx}
\usepackage{amssymb}
\newcommand{\gammap}{\dot{\gamma}}
\newcommand{\gammapA}{\dot{\gamma}_{\scriptscriptstyle A}}
\newcommand{\gammapB}{\dot{\gamma}_{\scriptscriptstyle B}}
\newcommand{\sigmaB}{\sigma_{\scriptscriptstyle B}}
\newcommand{\sigmaA}{\sigma_{\scriptscriptstyle A}}

\newcommand{\seuil}{\sigma_{\scriptscriptstyle 0}}

\newcommand{\gammapeff}{\gammap_{\hbox{\rm\scriptsize eff}}}
\newcommand{\sta}{R_{\scriptscriptstyle 2}}
\newcommand{\rot}{R_{\scriptscriptstyle 1}}

\begin{document}
\title{Shear-banding in a lyotropic lamellar phase\\ 
Part 1: Time-averaged velocity profiles}
\author{Jean-Baptiste Salmon}
\email{salmon@crpp-bordeaux.cnrs.fr}
\author{S\'ebastien Manneville}
\author{Annie Colin}
\affiliation{Centre de Recherche Paul Pascal, Avenue Schweitzer, 33600 PESSAC, FRANCE}
\date{\today}
\begin{abstract}
Using velocity profile measurements based on dynamic light scattering and coupled to 
structural and rheological measurements in a Couette cell, we present 
evidences for a shear-banding scenario in the shear flow of the {\it onion} texture of a 
lyotropic lamellar phase. {\it Time-averaged} measurements
clearly show the presence of structural shear-banding in the vicinity of a shear-induced transition, 
associated to the nucleation and growth of a highly sheared band in the flow. 
Our experiments also reveal the presence of slip at the walls of the Couette cell.
Using a simple mechanical approach, we demonstrate that our data confirms 
the classical assumption of the shear-banding picture, in which the interface between bands lies 
at a given stress $\sigma^\star$. We also outline the presence of large temporal fluctuations of the flow field,
which are the subject of the second part of this paper [Salmon {\it et al.}, submitted to Phys. Rev. E].
      
\end{abstract}
\maketitle

\section{Introduction \label{sec:introduction}}

A wide class of materials, referred to as {\it complex fluids}, exhibit  
a common feature: when submitted to a shear stress $\sigma$, their flow cannot
be described easily as in the case of simple fluids \cite{Larson:99}.
Indeed, the flow of simple liquids is entirely determined by the knowledge of
the viscosity $\eta$: when submitted to a shear stress $\sigma$ far from any hydrodynamic instability and after 
a very short transient, the velocity profile is linear and characterized by a shear rate $\gammap$,
given by the linear relation $\gammap = \sigma/\eta$ \cite{Guyon:94}.
In the case of complex fluids, such a linear relation between $\sigma$ and $\gammap$ 
does not hold any more. In polymeric fluids for instance, the flow tends to
decrease the {\it effective viscosity} $\eta\,\widehat{=}\,\sigma/\gammap$,
whereas disordered media like emulsions or foams can resist elastically
to a small applied stress but flow plastically under large $\sigma$ \cite{Larson:99}.
Such behaviors are due to the presence 
of a supramolecular architecture that leads to a coupling
between the structure of the fluid and the flow \cite{Edimbourg:00}.
The {\it shear-thinning} effect in polymeric fluids for instance, is due
to the alignment of the polymer chains along the flow direction, whereas the 
crossover between {\it pasty} and fluid states in glassy materials
is attributed to microscopic rearrangements of the structure \cite{Hebraud:97}. 

In some cases, because of a strong flow--structure coupling,
the flow can {\it induce} new organizations. In wormlike micellar systems for instance,
a shear flow can induce a nematic phase \cite{Berret:94,Schmitt:94}. In
lamellar phases, shear may induce structures that do no exist at rest, like
e.g. the {\it onion} texture \cite{Diat:93_2,Roux:93}. Such {\it Shear-Induced
Structures} (SIS) have been observed in a lot of complex fluids, especially
in lyotropic systems \cite{Hu_1:98,Eiser:00,Ramos:00}.
These behaviors are generally associated with a {\it spatial structuration}
of the fluid: the SIS nucleates at a critical stress or a critical shear rate
and expands in the flow as $\sigma$ (or $\gammap$) is increased.
Since the viscosity of the SIS differs from that of the original structure, 
the flow field is assumed to be composed of two macroscopic {\it bands} of different shear rates.  
This behavior, generically called {\it shear-banding}, seems to be a universal feature
of complex fluids. Its microscopic origin is not yet clearly understood: the shear-banding
behavior may arise from elastic instabilities \cite{Spenley:93}, from frozen disorder and local
plastic events \cite{Kabla:03} or from a coupling between shear flow and a phase transition
\cite{Olmsted:92}. Numerous phenomenological descriptions of shear-banding 
have emerged during the last decade 
(see Refs.~\cite{Lu:00,Goveas:01,Yuan:99,Ajdari:98,Picard:02} and references therein).
Those theoretical approaches have undoubtedly helped to understand the experiments, while not
always providing a microscopic description of the phenomenon.  

Most of experimental works relie on rheological data and structural measurements.
In those experiments, shear-banding is attributed to 
the presence of a {\it stress plateau} on the flow curve associated with 
a structural change, observed for instance using x-ray diffraction techniques \cite{Berret:94,Schmitt:94,Eiser:00}
or direct visualizations of birefringence bands in the flow field \cite{Cappelaere:97,Lerouge:98}.
Recently, a few experiments using Nuclear Magnetic Resonance velocimetry
have evidenced the fact that velocity profiles in such systems may display bands of different shear rates \cite{Mair:96,Britton:99}.
However, some of these experiments 
have reported a contradictory picture for the shear-banding instability: it seems that
the observed structural bands do not always correspond to bands of different shear rates, at least in a
specific wormlike micellar system \cite{Fischer:01}. 
More recently, experiments based on dynamic light scattering velocimetry performed in our group 
have revealed a classical shear-banding phenomenology in another wormlike micellar system \cite{Salmon:03_4}.
Therefore, more experimental data are needed to fully understand the spatial structuration of the flow field 
in the general SIS and shear-banding phenomenon.    

The aim of this paper is to perform an extensive study of shear-banding in a specific complex fluid,
a {\it lyotropic lamellar} phase. This system is known to exhibit a shear-induced transition, called the
{\it layering} transition, between two different {\it textures} of the phase \cite{Diat:95,Sierro:97}.
In the present article, we present {\it time-averaged} velocimetry measurements 
performed simultaneously to structural measurements
using respectively dynamic and static light scattering techniques.

This paper is organized as follows. In Sec.~\ref{sec:globalrheology}, we briefly describe the system under study 
and the experimental setup allowing us to measure both the structure of the fluid and its rheological behavior.
In Sec.~\ref{sec:experimentalsetup}, we present the experimental setup
for measuring velocity profiles in a Couette flow \cite{Salmon:03_2}.  
Sec.~\ref{sec:localrheology} describes the 
experiments performed at a given temperature ($T=30^\circ$C) and under controlled shear rate.
These experiments support the classical
picture of the shear-banding in this particular system and our data also reveals the presence of wall slip, which
is consistent with the observed shear-banding scenario.
A detailed mechanical analysis based on wall slip measurements and rheological data is presented in Sec.~\ref{asmmotsbi} and 
helps us demonstrate that the classical mechanical assumption of shear-banding, in which the interface between the bands
is only stable at a given stress, holds in our experiments. We then briefly summarize our results in Sec.~\ref{conclusion},
and emphasize on the puzzling presence of temporal fluctuations of the flow field near the shear-induced transition. 
The study of these temporal fluctuations are presented in a related paper where the link
with {\it rheochaos} is emphasized \cite{Salmon:03_5}.  

\section{Global rheology of the onion texture near the layering transition 
\label{sec:globalrheology}}
\subsection{Rheology of lyotropic lamellar phases:\\ the \textit{onion} texture}

Lyotropic systems are composed of surfactant molecules in a solvent. 
Because of the chemical duality 
of the surfactant molecules (polar head and long hydrophobic chain), 
self-assembled structures at the nanometer scale are commonly observed \cite{Larson:99}.
 Depending on the range of concentrations, one can observe different structures: 
for instance, long cylinders of some microns long ({\it wormlike micelles}) or infinite 
surfactant bilayers ({\it membranes}). In the latter case, because of the interactions,
the membranes can form periodic stacks, referred to as \textit{lamellar} phases 
and noted $L_{\alpha}$, or randomly 
connected continuous structures, called \textit{sponge} phases and 
noted $L_{\scriptscriptstyle 3}$ \cite{Porte:91,Roux:92}. 
In a lamellar phase, the smectic period $d$, i.e. the distance 
between the bilayers,
ranges from a few nanometers up to 1~$\mu$m whereas the thickness $\delta$ 
of the membranes is 2--4~nm. 
These liquid crystalline phase are locally organized as a monocrystal of smectic phase. If no specific
treatment is applied, lyotropic lamellar phases contain at larger length scales ($\approx 10~\mu$m) a 
lot of characteristic defects: the organization of these 
defects is called the \textit{texture} of the phase.    

Since the work of Roux and coworkers, great experimental effort 
has been devoted to the understanding of the effect of a shear flow on lamellar 
phases \cite{Diat:93_2,Roux:93}. A robust experimental fact has emerged: the shear flow controls
the texture of the phase. 
For most systems, the experimental behavior usually observed is as follows. 
(i) At very low shear rates
($\gammap \leq 1$~s$^{-1}$), the membranes tend to align with the direction of 
the flow but textural defects still persist. (ii) Above a critical shear rate of about 
$\gammap\approx 1$~s$^{-1}$, the membranes are wrapped into monodisperse multilamellar vesicles called \textit{onions} 
and organized as a disordered close-compact texture. 
The characteristic size $R$ of this shear-induced structure
(the onion size) is of the order of a few microns. 
(iii) At higher shear rates ($\approx 10^3$~s$^{-1}$), perfectly ordered lamellar 
phases are recovered \cite{Diat:93_2}.
For some specific systems and at intermediate shear rates, spatial organizations of onions are 
sometimes encountered \cite{Diat:95,Sierro:97}. 

The complex fluid investigated here corresponds to this last behavior. It is made of Sodium Dodecyl Sulfate (SDS) 
and Octanol (surfactant molecules) in Brine (solvent). At the concentrations considered here 
(6.5\%~wt. SDS, 7.8\%~wt. Octanol, and 85.7\%~wt. Brine at 20 g.L$^{-1}$), a lamellar 
phase is observed \cite{Herve:93}. The smectic period $d$ is 15~nm and the bilayers thickness $\delta$ 
is about  2~nm. For the given range of concentrations, the lamellar 
phase is stabilized by undulating interactions \cite{Helfrich:78}. This system is 
very sensitive to temperature: 
for $T\geq T_{{L_{\alpha}}\hbox{--}{L_{\scriptscriptstyle 3}}} \approx 35^\circ$C, 
a sponge--lamellar phase mixture appears.

\begin{figure}[ht]
\begin{center}
\scalebox{1.0}{\includegraphics{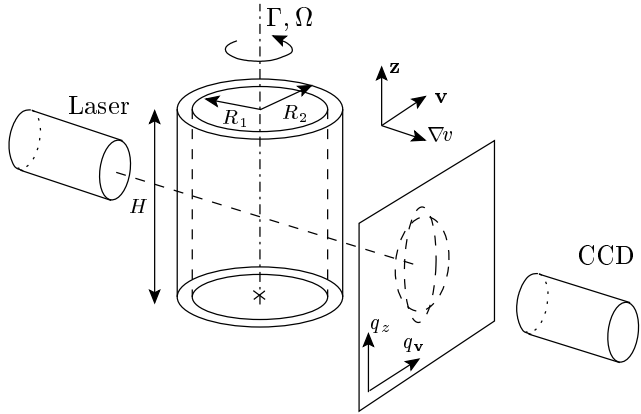}}
\end{center}
\caption{Experimental setup. A thermostated plate (not shown) on top of the cell 
allows us to avoid evaporation. The rheometer imposes a constant torque $\Gamma$ 
on the axis of the Couette cell and records its rotation speed $\Omega$. The geometry of the Couette cell is:
$H=30$~mm,  $\rot=24$~mm, and $\sta=25$~mm, leading to a gap width $e=\sta-\rot=1$~mm.}
\label{setup}
\end{figure}

To study the effect of shear flow on this phase, we used the experimental device sketched 
in Fig.~\ref{setup}. A rheometer (TA Instruments AR1000N) and a 
Couette cell made of Plexiglas (gap $e=1$~mm, inner radius $\rot= 24$~mm and height $H=30$~mm)
allow us to perform rheological measurements.
The rheometer imposes a constant torque $\Gamma$ on the axis of the Couette cell
which induces a constant stress $\sigma$ in the fluid. 
The rotation speed $\Omega$ of the Couette cell
is continuously recorded, from which the shear rate $\gammap$ can be deduced.
A computer-controlled feedback loop on the applied torque $\Gamma$, 
can also be used to apply a constant shear rate without any significant 
temporal fluctuations ($\delta\gammap/\gammap \approx 0.01\%$). 
The relations between ($\sigma$,$\gammap$) given by the rheometer 
and ($\Gamma$,$\Omega$) read:
\begin{eqnarray}
\sigma &=& \frac{R_{\scriptscriptstyle 1}^2+R_{\scriptscriptstyle 2}^2}
{4\pi H R_{\scriptscriptstyle 1}^2 R_{\scriptscriptstyle 2}^2}\,\Gamma\,,\label{e.sigmarheo} \\ 
\gammap &=& \frac{R_{\scriptscriptstyle 1}^2+R_{\scriptscriptstyle 2}^2}
{R_{\scriptscriptstyle 2}^2-R_{\scriptscriptstyle 1}^2}\,\Omega\, .
\label{e.gammarheo}
\end{eqnarray}
Such definitions ensure that ($\sigma$,$\gammap$)  correspond 
to the average values of the local stress and shear rate in the case of a Newtonian fluid.
Temperature is controlled within $\pm $0.1$^\circ$C 
using a water circulation around the cell.
A polarized laser beam (He--Ne, $\lambda = 632.8$~nm) is directed through the 
transparent Couette cell to investigate the effect of a shear flow 
on the structure of the fluid. Since
the characteristic sizes of the texture are about 1~$\mu$m, well-defined
diffraction patterns are obtained and recorded on a Charge-Coupled Device camera (CCD, Cohu).  
As shown in Fig.~\ref{setup}, the laser beam crosses the sample twice, so that  
two diffraction patterns are recorded; the first one has an elliptical shape due to 
the optical refractions induced by the Couette geometry.   

\subsection{The layering transition: stationary state and rheological chaos \label{ssec:tlt:ssarc}}

\begin{figure}[htbp]
\begin{center}
\scalebox{1.0}{\includegraphics{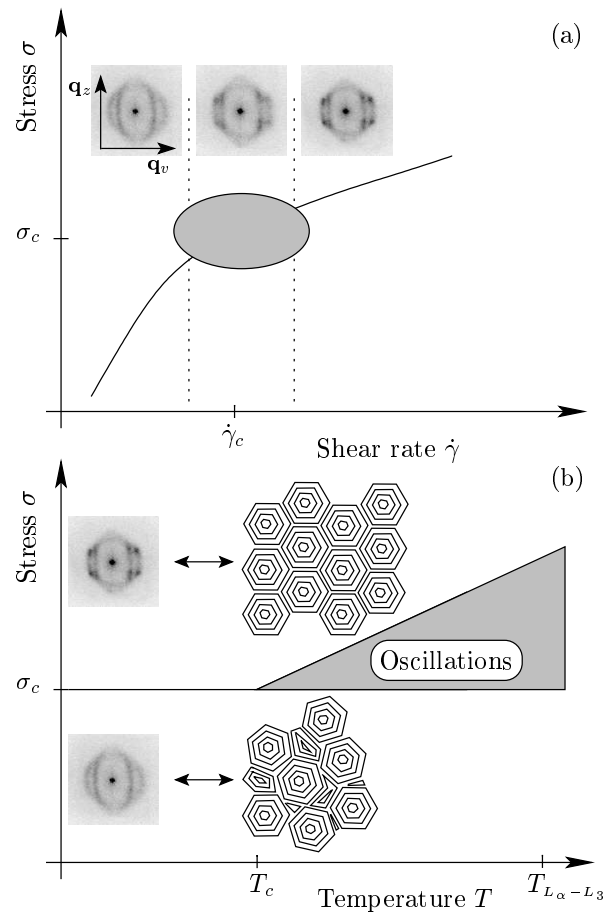}}
\end{center}
\caption{(a) Typical flow curve of the onion texture. The diffraction pattern corresponds to the structure under shear.
The gray area indicates the region of the oscillations of the 
shear rate at imposed stress for $T \geq T_c \approx 27^\circ$C.  
(b) Shear diagram $\sigma$ vs. $T$. Oscillations of the shear rate at imposed stress appear for temperature 
$T \geq T_c \approx 27^\circ$C. Shown are
the schematic representations of the disordered and ordered states.
$T_{{L_{\alpha}} \hbox{--} {L_{\scriptscriptstyle 3}}} \approx 35^\circ$C in the system under study.}
\label{schematic1}
\end{figure}

Figure~\ref{schematic1}(a) shows a typical flow curve obtained on our lamellar phase at
imposed stress or imposed shear rate for $T<T_c \approx 27^\circ$C.
After a transient phase (at least 10--30 min), a stationary state is obtained, 
i.e. $\sigma$ ($\gammap$ resp.) does not vary significantly in time when $\gammap$
($\sigma$ resp.) is imposed. At low stress, the diffraction
patterns are uniform rings indicating the presence of a disordered texture of onions. 
When $\sigma$ reaches $\sigma_c$ ($\approx 15$~Pa), six peaks
appear on the ring indicating that the onions get a long range hexagonal order on layers
sliding onto each other (see Fig.~\ref{schematic1}) \cite{Diat:95,Sierro:97}. 
Those layers lie in the ($\mathbf{v},\mathbf{z}$) plane normally 
to the shear gradient direction $\nabla\!v$. When $\sigma$ is  
increased further, peaks become more contrasted. 
Note that the peaks with wave-vectors along the rheometer axis 
($\mathbf{q}_z$) are less intense than the others. 
This is due to the zig-zag motion of the planes of onions when sliding 
onto each other \cite{Sierro:97}.
This shear-induced ordering transition, called the \textit{layering} 
transition, has been observed in many monodisperse colloidal systems since 
the pioneering work of Ackerson and Pusey \cite{Ackerson:88}. This ordering transition is 
associated to a shear-thinning of the fluid, i.e. the ordering of the 
colloids causes the decrease of the viscosity.

More recently, complex dynamical behaviors have been observed 
in the vicinity of the layering transition for temperatures $T>T_c \approx 27^\circ$C 
(see Fig.~\ref{schematic1})
and under controlled stress: 
the shear rate does not reach a stationary value, 
but sometimes oscillates indefinitely in time \cite{Wunenburger:01,Salmon:02}.
However, when the shear rate is imposed, no oscillations
are observed and the responses of the shear stress seem almost stationary. 
In the stress imposed case, the shear rate oscillations are characterized by a 
large period of about 10 min.
Moreover, the dynamics of the shear rate is correlated to a dynamical structural change: the 
fluid oscillates between the disordered and the layered states. 
The origin of such surprising dynamics is not yet understood, but it is now quite clear 
that it does not correspond to a simple hydrodynamic or elastic instability. Such a
new kind of temporal instability was coined \textit{rheochaos} \cite{Grosso:01,Cates:02}.
In a previous work \cite{Salmon:02}, a detailed study of such dynamics 
was performed: depending on the 
applied stress, several dynamical regimes have been found. 
Using dynamical system theory, a careful analysis 
has revealed that the dynamics do not simply correspond to 
a low-dimensional chaotic system, probably because
some spatial degrees of freedom are involved. Moreover, 
rheological experiments performed with different gap widths $e$ 
support the assumption of a spatial organization of the flow in the gap \cite{Salmon:02}. 
Our idea was that spatial structures such as \textit{bands}, that lie 
in the $\mathbf{\nabla}\!v$ direction and
oscillates in time, could lead to the observed dynamics. 
Therefore, it was essential to measure the {\it local} velocity
rather than the {\it global} rotation speed $\Omega$ of the Couette cell.

In the next section, we present the experimental setup allowing 
us to perform velocimetry measurements. 
We then focus on experiments performed on 
{\it rheological stationary states} at $T=30^\circ$C and under controlled shear rate.

\section{Experimental setup: local velocimetry using Dynamic Light 
Scattering (DLS) \label{sec:experimentalsetup}}

\begin{figure}[htbp]
\begin{center}
\scalebox{1.0}{\includegraphics{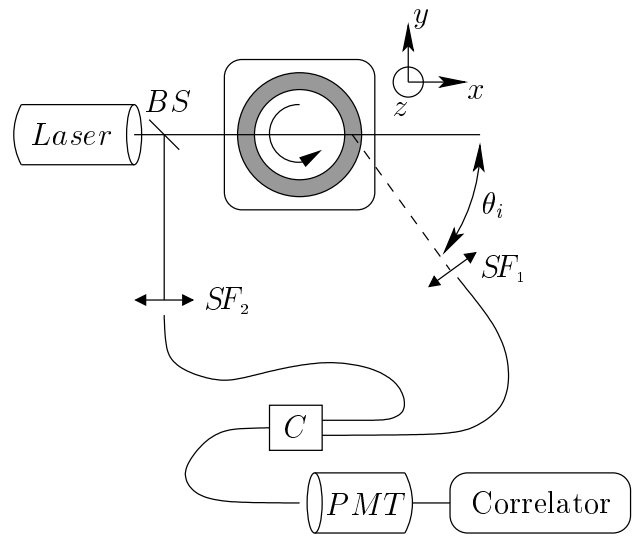}}
\end{center}
\caption{Heterodyne DLS setup. $BS$ denotes a beam splitter, 
$SF$ spatial filters, and $C$ the device
coupling optical fibers and used to perform the interference 
between the scattered light and the reference beam.}
\label{setup2}
\end{figure}

In order to evidence the possibility of a spatial organization of the fluid flow 
near the layering transition,
we have used the experimental setup sketched in Fig.~\ref{setup2} 
and described in more details in Ref.~\cite{Salmon:03_2}. 
It consists of a classical heterodyne Dynamic Light Scattering experiment (DLS)  
mounted around a rheometer. 
Local velocity measurements using heterodyne DLS rely on the detection of the 
Doppler frequency shift
associated with the motion of the scatterers inside a small scattering volume $\mathcal{V}$
\cite{Berne:95,Ackerson:81,Fuller:80,Maloy:92,Gollub:74}.
In classical heterodyne setups, light scattered by the sample under 
study is collected along 
a direction $\theta_i$ and is made to interfere with a reference beam. 
Light resulting from the interference is sent to a photomultiplier tube (PMT)
and the auto-correlation function $C(\tau)$ of the intensity is computed
using an electronic correlator. 

\begin{figure}[htbp]
\begin{center}
\scalebox{1}{\includegraphics{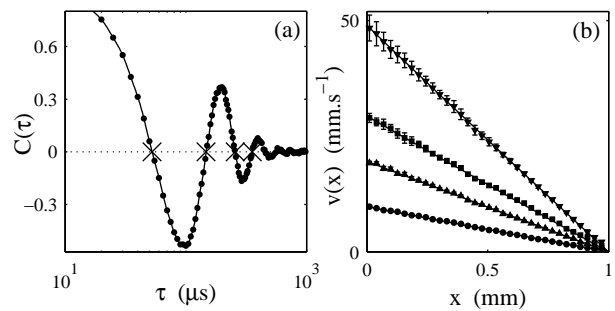}}
\end{center}
\caption{(a)  Experimental heterodyne correlation function ($\bullet$) recorded on a latex 
suspension at $\gammap=10$~s$^{-1}$.
The solid line corresponds to an interpolation of the heterodyne
function from which the frequency shift $\mathbf{q}\cdot\mathbf{v}$ is computed. 
(b) Different velocity profiles obtained
for various applied shear rates, 
$\gammap=10$ ($\bullet$), 
$20$ ($\scriptscriptstyle{\blacktriangle}$), 
$30$ ($\scriptscriptstyle{\blacksquare}$), 
and $50$~s$^{-1}$ ($\blacktriangledown$). The solid lines are 
the expected velocity profiles for a Newtonian fluid.}
\label{exphete}
\end{figure}

When the scattering volume $\mathcal{V}$ is submitted to a shear flow,
it can be shown that the correlation function $C(\tau)$ is an oscillating
function of the time lag $\tau$ modulated by a slowly decreasing envelope.
The frequency of the oscillations in $C(\tau)$
is exactly the Doppler shift $\mathbf{q}\cdot\mathbf{v}$, where $\mathbf{q}$
is the scattering wavevector and $\mathbf{v}$ is the local velocity averaged
over the size of the scattering volume $\mathcal{V}$. 
In the experiments presented here, the imposed angle $\theta_i=62.5^\circ$
has been chosen in such a way that 
the size of $\mathcal{V}$ is about about 50~$\mu$m \cite{Salmon:03_2}.
Figure~\ref{exphete}(a) shows a typical correlation function 
measured on a sheared latex suspension.
The frequency shift $\mathbf{q}\cdot\mathbf{v}$ is recovered by interpolating a
portion of $C(\tau)$ and looking for the zero crossings. Errorbars on such measurements
are obtained through the fitting procedure: typical uncertainties are less than 5~\%.

The rheometer sits on a mechanical table whose displacements are controlled by a computer.
Three mechanical actuators allow us to move the rheometer in the $x$, $y$, and
$z$ directions with a precision of $1~\mu$m. Once $y$ is set so that the incident
beam is normal to the cell surface, velocity profiles are measured by moving the mechanical
table in the $x$ direction by steps of 30~$\mu$m. 

As discussed in Ref.~\cite{Salmon:03_2}, going from $\mathbf{q}\cdot\mathbf{v}$ 
as a function
of the table position to the velocity profile $v(x)$ 
requires a careful calibration procedure to take into account the
refraction effects due to the Couette 
cell. Figure~\ref{exphete}(b) presents velocity profiles measured at various known shear rates
during the calibration using a Newtonian suspension of latex spheres
in a water--glycerol mixture. This simple liquid has the same optical index ($n=1.35$)
as the lamellar phase under study.
Finally, we accumulate the correlation functions over 
3--5~s so that a full velocity profile
with a resolution of 30~$\mu$m takes about 3~min to complete. 

\section{Local rheology of the onion texture near the layering transition}
\label{sec:localrheology}

\subsection{Global rheology and local velocimetry \label{gralv}}

\begin{figure}[htbp]
\begin{center}
\scalebox{1}{\includegraphics{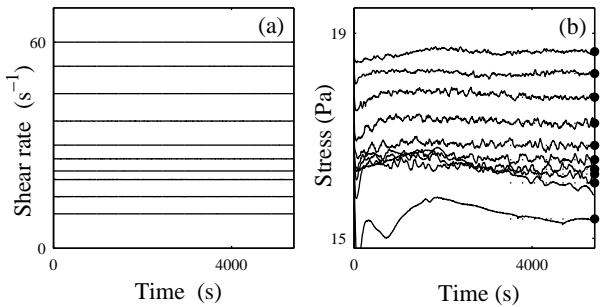}}
\end{center}
\caption{$T=30^\circ$C.
(a) Applied steps of shear rate. No significant temporal fluctuations are observed.
(b) Corresponding temporal responses of $\sigma(t)$. 
The dotted lines indicate the 
part of the time series where the average stress ($\bullet$) is computed. 
\label{reponse_rheol_T30}}
\end{figure}

To understand the coupling between spatial degrees of freedom, global rheology and
the structural transition, we first 
present measurements obtained at $T=30^\circ$C 
in the shear rate imposed mode.  
Well-defined rheological stationary states are obtained with such 
a method (see Sec.~\ref{ssec:tlt:ssarc}).
We record global rheological data simultaneously to the velocity 
profile measurements. In order to obtain 
reproducible experiments, careful rheological protocols must be 
used as discussed in Ref.~\cite{Salmon:02}.
At the temperature under study ($T=30^\circ$C), we apply
a first step at $\gammap=5$~s$^{-1}$ during 7200~s.
This step of applied shear rate allows us to begin the experiment 
with a well-defined stationary 
state of disordered onions. We then apply increasing shear 
rates for 5400~s per step. The increment
between two steps is $\delta\gammap = 2.5$--5~s$^{-1}$. 
The smaller value of $\delta\gammap$ is chosen when a fine resolution is needed
in the vicinity of the transition.
As shown in Fig.~\ref{reponse_rheol_T30}(b), the temporal responses of the shear 
stress are almost stationary: 
small temporal fluctuations ($ \delta\sigma / \sigma \approx 1$--3\%) can be detected. 
Note that, in the shear rate imposed mode the temporal fluctuations 
of $\gammap$ 
are completely negligible ($\delta\gammap/\gammap \approx 0.01\%$, 
[see Fig.~\ref{reponse_rheol_T30}(a)]).

\begin{figure}[htbp]
\begin{center}
\scalebox{1}{\includegraphics{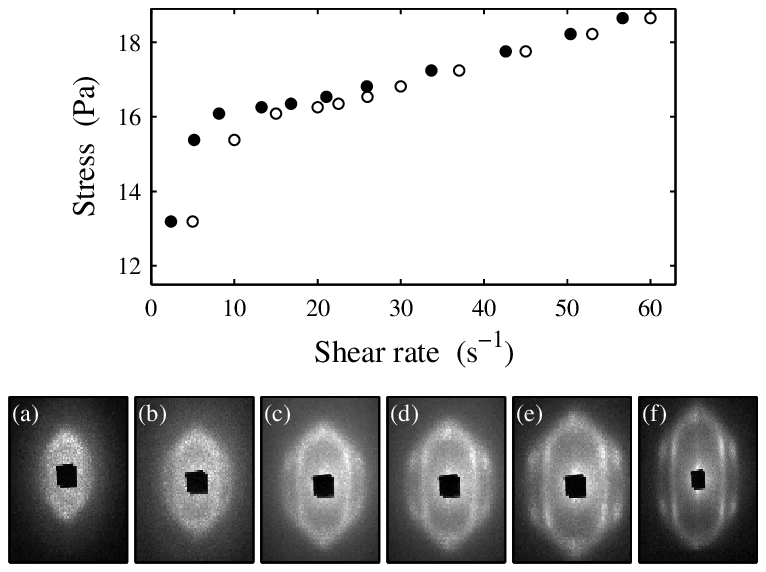}}
\end{center}
\caption{($\circ$) Stationary flow curve obtained at $T=30^\circ$C 
under controlled shear rate.
The values ($\sigma$,$\gammap$) are extracted from the temporal responses 
displayed in Fig.~\ref{reponse_rheol_T30}.
($\bullet$) Effective flow curve $\sigma$ vs. $\gammap_{\hbox{{\tiny eff}}}$. 
The effective shear rates take into account the effect of wall slip and 
are calculated using Eq.~(\ref{e.effectiveshear}) (see Sec.~\ref{ss:wsefcall}).
Corresponding diffraction patterns at 
$\gammap=10$ (a),
$15$ (b),
$22.5$ (c),
$26$ (d),
$30$ (e), and
$45$~s$^{-1}$ (f).
The field of the CCD camera has been adapted to the pattern size in the last diffraction pattern.}
\label{t30rheol}
\end{figure}

Global rheological measurements are displayed in Fig.~\ref{t30rheol} and 
the corresponding diffraction patterns at various applied
shear rates are shown in Figs.~\ref{t30rheol}(a)--(f). 
For shear rates below $\gammap_c \approx 15$~s$^{-1}$, 
diffraction patterns are uniform rings indicating that the structure of the 
onions is disordered [Figs.~\ref{t30rheol}(a)-(b)]. 
Above $\gammap_c$, six fuzzy peaks appear on the diffraction ring,
indicating the onset of the layering transition [Fig.~\ref{t30rheol}(c)]. 
When the shear rate is further increased, peaks become more 
and more contrasted as can be seen in Figs.~\ref{t30rheol}(d)--(f).
The flow curve does not display any significant discontinuity at the layering transition: 
it is rather difficult to locate the various regions of different structures from 
the rheological data alone and without the information inferred from the diffraction patterns. 
Significant shear-thinning is observed: $\eta \approx 2.6$~Pa.s at $\gammap = 5$~s$^{-1}$ 
and  $\eta \approx 0.3$~Pa.s at $\gammap = 60$~s$^{-1}$.

\begin{figure}[htbp]
\begin{center}
\scalebox{1}{\includegraphics{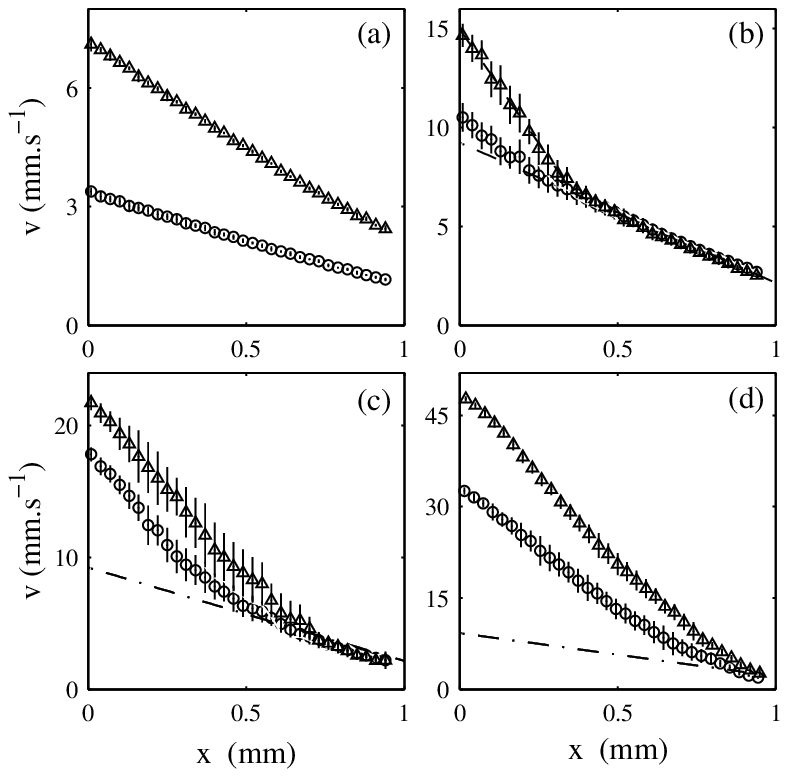}}
\end{center}
\caption{Time-averaged velocity profiles obtained simultaneously to the 
flow curve displayed in Fig.~\ref{t30rheol}.
The errorbars represent the temporal fluctuations of the local velocity (see text). 
(a) 
$\gammap=5$ ($\circ$) and
$10$~s$^{-1}$ ($\triangle$).
(b) 
$\gammap=15$ ($\circ$), and
$20$~s$^{-1}$ ($\triangle$).
The dashed line is a guide for the eye in the highly sheared band and corresponds 
to a local shear rate $\gammapB\approx23$~s$^{-1}$.
(c) 
$\gammap=22.5$ ($\circ$) and 
$26$~s$^{-1}$ ($\triangle$).
(d)
$\gammap=37$ ($\circ$) and
$53$~s$^{-1}$ ($\triangle$).
For (b)--(d), the dash-dotted line indicates the weakly sheared band and corresponds to 
a local shear rate $\gammapA \approx 7.1$~s$^{-1}$.
\label{profilt30}}
\end{figure}

Far below or far above the layering transition, 
well-defined profiles are easily measured. 
But very surprisingly, near the layering transition, the local velocity is not stationary but 
displays large temporal fluctuations while $\sigma$ only fluctuates within a few percents 
(see Ref.~\cite{Salmon:03_5} for a complete study). 
The characteristic times of these fluctuations 
range from 100 to 1000~s and are of the order of the time needed 
to obtain a full velocity profile (2--3~min). Such
dynamics prevent us to record well-defined profiles. 
To obtain valuable data, we chose to measure {\it time-averaged} 
velocity profiles: several 
profiles were recorded at a given applied shear rate, and later averaged. 
The typical number of profiles needed to obtain
good statistical estimates of the flow field ranges 
between 10 and 20. The standard deviations of 
those estimates yield the amplitudes of the temporal fluctuations.     
Figure~\ref{profilt30} displays velocity profiles measured simultaneously to 
the flow curve of Fig.~\ref{t30rheol}.
Each profile corresponds to an average over up to 20 measurements. 
Errorbars represent the standard deviations of the local velocities.
The main features of these profiles are as follows.
(i) Below the layering transition i.e. for 
$\gammap\leq\gammap_{\scriptscriptstyle 1}\approx 15$~s$^{-1}$, 
velocity profiles are perfectly stationary. The flow is homogeneous since the 
profiles are nearly linear. However, significant wall slip can be detected: the velocity at the stator does not
vanish at $x=e$, and the fluid velocity does not reach the rotor velocity at $x=0$.
(ii) At $\gammap \gtrsim \gammap_{\scriptscriptstyle 1}$, a 
highly sheared band nucleates at the rotor. 
The shear rate in this band is about $\gammapB\approx23$~s$^{-1}$ and 
the other band is sheared at $\gammapA\approx7.1$~s$^{-1}$ [see Fig.~\ref{profilt30}(b)]. 
Significant temporal fluctuations are present \cite{Salmon:03_5}, and important wall slip can still be measured.
(iii) When the shear rate is further increased up to 
$\gammap=\gammap_{\scriptscriptstyle 2}\approx37$~s$^{-1}$, 
the highly sheared band covers the whole gap. The value 
$\gammapA\approx7.1$~s$^{-1}$ of the weakly sheared band remains almost constant 
over the coexistence domain (15--37~s$^{-1}$).
(iv)  For $\gammap \geq \gammap_{\scriptscriptstyle 2}$, 
the flow becomes homogeneous again. Temporal fluctuations of the velocity also 
disappear but the fluid still slips at the two walls.    

Those results show without any ambiguity that the layering 
transition can be described by the classical phenomenology of {\it shear-banding}.
At $\gammap\leq\gammap_{\scriptscriptstyle 1}$, the flow is 
homogeneous and corresponds to a disordered state of onions,
as can be checked from the diffraction patterns.
At $\gammap\gtrsim\gammap_{\scriptscriptstyle 1}$, velocity profiles 
display two bands corresponding to two given shear rates $\gammapA$ and $\gammapB$.
At $\gammap\gtrsim\gammap_{\scriptscriptstyle 1}$, peaks 
also appear on the ring. 
When the shear rate is further increased up to $\gammap_{\scriptscriptstyle 2}$, 
the width of the highly sheared band grows as well as the 
constrast of the peaks on the diffraction ring. 
At $\gammap\geq\gammap_{\scriptscriptstyle 2}$, the flow is 
homogeneous again and the contrast of the peaks on
the ring is maximal.

Let us recall that the diffraction patterns correspond to
a measure of the structure of the fluid {\it integrated} along the 
velocity gradient direction $\nabla\!v$ (see Fig.~\ref{setup}).
Therefore, the contrast of the peaks on the ring provides an estimate of the relative proportion of layered vs. disordered onions
in the gap. Thus, the above results clearly point to a picture of the flow where the highly (weakly resp.) sheared band 
corresponds to the layered (disordered resp.) onions. To our knowledge, this data set brings the first experimental
evidence for both structural shear-banding and banded flows.    

In the next paragraph, we present a detailed analysis of the slip velocities. 
We also demonstrate that the correction due to wall slip allows us to use 
a mechanical model to fit the velocity profiles consistently with the classical picture of shear-banding.     

\subsection{Wall slip, effective flow curve and lubricating layers \label{ss:wsefcall}}

\subsubsection{Effective flow curve}
Using the measured velocity profiles,
it is easy to define an {\it effective} shear rate in the bulk
onion texture by $\gammapeff=(v_{\scriptscriptstyle 1}-v_{\scriptscriptstyle 2})/e$, where 
$v_{\scriptscriptstyle 1}$ ($v_{\scriptscriptstyle 2}$ resp.) corresponds to the velocity of the fluid
at $x=0$ ($x=e$ resp.) estimated from the measured velocity profiles. 
Moreover, in order to take into account the stress inhomogeneity in the Couette cell and
thus get a quantitative comparison between this effective shear rate and the global
shear rate $\gammap$ indicated by the rheometer, we define $\gammapeff$ consistently
with Eq.~(\ref{e.gammarheo}) by:
\begin{equation}
\gammapeff=\frac{R_{\scriptscriptstyle 1}^2+
R_{\scriptscriptstyle 2}^2}{R_{\scriptscriptstyle 1}(R_{\scriptscriptstyle 1}
+R_{\scriptscriptstyle 2})}\,\frac{v_{\scriptscriptstyle 1}-v_{\scriptscriptstyle 2}}{e}\,.
\label{e.effectiveshear}
\end{equation}
Figure~\ref{t30rheol} presents the {\it effective} flow curve $\sigma$($\gammapeff$) ($\bullet$)
and the flow curve computed from the values indicated by the rheometer 
$\sigma$($\gammap$) ($\circ$). The effective shear rate allows us to remove the contribution
due to wall slip. When compared to the global flow curve $\sigma$ vs. $\gammap$, 
the effective flow curve seems to reveal a stress plateau
at a value of about $16$~Pa. This plateau extends 
from $\gammapA\approx 7$~s$^{-1}$ to $\gammapB\approx 25$~s$^{-1}$.
Note that this stress plateau is not perfectly {\it flat} but
presents a small slope. We will show in Sec.~\ref{asmmotsbi} that such a slope
is due to the curvature of the Couette geometry as expected from theoretical models \cite{Radulescu:00}.

\subsubsection{Lubricating layers}

To explain the presence of wall slip in complex fluid flows, 
one usually considers that thin lubricating layers
are present at the walls of the Couette cell. In emulsions 
for instance, it is well established
that wall slip is due to the presence of highly sheared thin films composed 
of the continuous phase only 
\cite{Barnes:95}. These films play the role of {\it lubricating layers}: 
the bulk material is weakly 
sheared, whereas the films {\it absorb} a part of the viscous stress.
In our case, we can reasonably assume that wall slip is due to 
very thin layers composed of water or of a 
few membranes lying normally to the velocity gradient 
direction $\nabla\!v$.
The resolution of our setup ($\approx 50~\mu$m) 
does not allow us to measure {\it directly} the thicknesses of those layers. 

However, in this simple picture, there is no discontinuity 
of the shear stress inside the gap of the Couette cell.
Because the thicknesses of the lubricating layers are very small
($\approx 100$~nm), one can assume that the flow inside the films is  
laminar. Under this assumption, we may access the 
thicknesses of the lubricating layers.
Indeed, let us define the slip velocity at the rotor 
$v_{s_{\scriptscriptstyle 1}}=v_{\scriptscriptstyle 0}-v_{\scriptscriptstyle 1}$ 
as the difference between the rotor velocity 
$v_{\scriptscriptstyle 0}$ and the velocity $v_{\scriptscriptstyle 1}$ 
at the rotor, and the slip velocity $v_{s_2}=v_{\scriptscriptstyle 2}$ at the stator
as the velocity $v_{\scriptscriptstyle 2}$ measured at the stator.
$v_{si}/h_i$ then corresponds to the mean shear rate inside the film of thickness $h_i$
($i=1$ for the rotor and $i=2$ for the stator).
By assuming that the stress is continuous inside the gap of 
the Couette cell, the thicknesses of the layers are then given by:
\begin{equation}
h_i=\frac{\eta_{f} v_{s_i}}{\sigma_i}\,, 
\label{hfilm}
\end{equation}
where $\sigma_i$ is the shear stress near wall number $i$ and $\eta_f$ is
the viscosity of the lubricating layers. 
Note that in the Couette geometry the local stress is given by $\sigma(r)=\sigma_{\scriptscriptstyle 1}\rot^2/r^2$, 
where $\sigma_{\scriptscriptstyle 1}$ is the stress at the rotor.
The values $\sigma_i$ at the walls are linked to $\sigma$ according to:
\begin{equation}
\sigma_i=\frac{2R_j^2}{R_{\scriptscriptstyle 1}^2+R_{\scriptscriptstyle 2}^2}\,\sigma\,,
\label{e.sigmawall}
\end{equation}
where $j=2$ (resp. $j=1$) when $i=1$ (resp. $i=2$) and $\sigma$ is the global value indicated by the rheometer 
[see Eq.~(\ref{e.sigmarheo})].

\subsubsection{Wall slip and shear-banding}

\begin{figure}[htbp]
\begin{center}
\scalebox{1}{\includegraphics{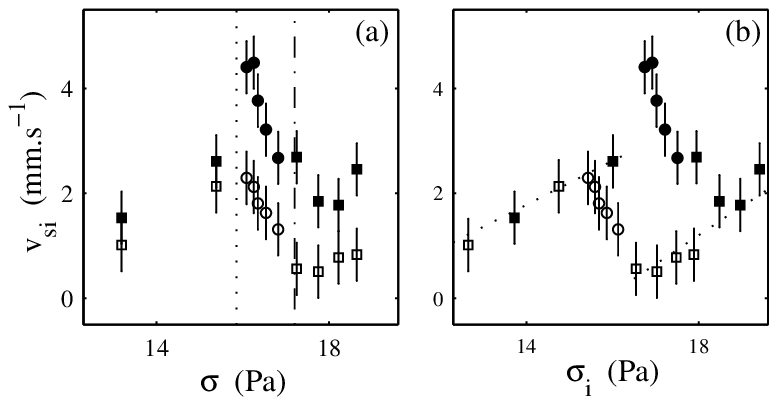}}
\end{center}
\caption{(a) Sliding velocities  $v_{si}$ vs. stress $\sigma$:
$v_{s{\scriptscriptstyle 1}}$
at the rotor ($\blacksquare$,$\bullet$) and 
$v_{s{\scriptscriptstyle 2}}$ 
at the stator ($\square$,$\circ$).
The dotted line indicates $\sigmaA=15.9$~Pa and the 
dash-dotted line $\sigmaB=17.2$~Pa (see text).
(b) $v_{si}$ vs. local stress $\sigma_i$. The dotted lines are guide lines for the 
homogeneous states.
($\square$,$\blacksquare$) correspond to homogeneous velocity profiles and
($\circ$,$\bullet$) to banded velocity profiles. 
\label{vs_sigma_T30}}
\end{figure}

Figure~\ref{vs_sigma_T30}(a) presents the slip velocities $v_{si}$
vs. the global stress $\sigma$, both at the stator 
($\square$,$\circ$) and at the rotor ($\blacksquare$,$\bullet$).
The homogeneous profiles are indicated by squares ($\square$,$\blacksquare$)
and banded flows by circles ($\circ$,$\bullet$). 
Figure~\ref{vs_sigma_T30}(a) reveals the three regimes previously observed
on both velocity profiles and diffraction patterns:
(i) when $\sigma<\sigmaA\approx 15.9$~Pa, the velocity profiles are nearly linear
and the flow is only composed of disordered onions. In this region,
the slip velocities $v_{si}$ slightly increase under increasing stress $\sigma$.   
(ii) When $\sigmaA\leq\sigma\leq\sigmaB\approx 17.2$~Pa, the nucleation and growth of the 
highly sheared band corresponding to the layered state is associated to 
a decrease of $v_{si}$ with $\sigma$. (iii) When $\sigma>\sigmaB$, the flow 
is homogeneous again and the slip velocities $v_{si}$ slightly 
increase with $\sigma$. Note that the exact values of $\sigmaA$ and $\sigmaB$
indicating the coexistence domain will be calculated using Eqs.~(\ref{siga_sigb1}) and (\ref{siga_sigb2}) in
Sec.~\ref{asmmotsbi}.

Figure~\ref{vs_sigma_T30}(b) presents the same slip velocities but plotted against
the local stress $\sigma_i$ inferred from Eq.~(\ref{e.sigmawall}).
It reveals two important results.
(i) When the flow is homogeneous, the slip velocities
at the rotor $v_{s\scriptscriptstyle 1}$ 
and at the stator $v_{s\scriptscriptstyle 2}$ collapse on the same curve
when plotted against the local stresses $\sigma_{\scriptscriptstyle 1}$ and
$\sigma_{\scriptscriptstyle 2}$.
In other words, $v_{si}$ seem to be a unique function 
of the local stress in the homogeneous domains, i.e. $v_{si} = f(\sigma_i)$.
This feature is also observed in concentrated colloidal systems where inertial effects 
are totally supressed by the osmotic pressure needed to concentrate the colloids 
\cite{Salmon:03_1,Barnes:95}. 
(ii) In the coexistence domain, i.e. for $\sigmaA\leq\sigma\leq\sigmaB$,   
there is a large difference between slip velocities at the rotor and at the stator,
even when $v_{si}$ are plotted against the local stress $\sigma_i$.
This result confirms
the observed structuration: wall slip is very different at the two walls
because there are {\it two} different fluids inside the gap (the layered state 
of onions lies near the rotor whereas the disordered onions lie 
near the stator). In the coexistence domain, the fact that the slip velocity $v_{si}$
is a unique function of the local stress $\sigma_i$ does not hold anymore.  
Moreover, our data clearly indicate that 
wall slip is larger at the rotor than at the stator in this domain. 

\subsubsection{Thicknesses of the lubricating layers}

\begin{figure}[htbp]
\begin{center}
\scalebox{1}{\includegraphics{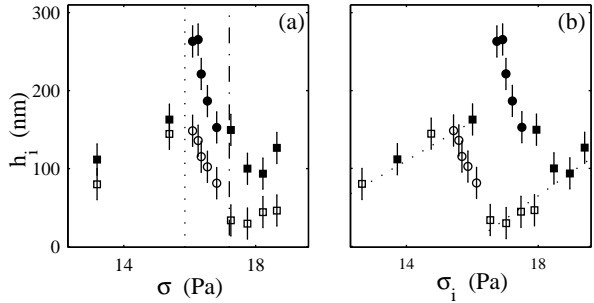}}
\end{center}
\caption{(a) Thicknesses $h_i$ of the sliding layer vs. stress $\sigma$: 
$h_{{\scriptscriptstyle 1}}$ 
at the rotor ($\blacksquare$,$\bullet$) and 
$h_{{\scriptscriptstyle 2}}$ 
at the stator ($\square$,$\circ$).
The dotted line indicates $\sigmaA=15.9$~Pa 
and the dash-dotted line $\sigmaB=17.2$~Pa (see text).
(b) $h_{i}$ vs. local stress $\sigma_i$. The dotted lines are guide lines for the 
homogeneous states.
($\square$,$\blacksquare$) correspond to homogeneous velocity profiles and
($\circ$,$\bullet$) to banded velocity profiles. }\label{hs_sigma_T30}
\end{figure}

Figure~\ref{hs_sigma_T30} presents the thicknesses $h_i$ of the lubricating
layers calculated from Eq.~(\ref{hfilm}) and 
assuming that the viscosity of the lubricating films is $\eta_f=10^{-3}$~Pa.s, the 
viscosity of water. The thicknesses $h_i$ are of the order of 100--200~nm. Such a value
is in good agreement with other measurements performed in concentrated colloidal systems
\cite{Barnes:95,Salmon:03_1}. Let us recall that in the case of our 
lamellar phase, the smectic period $d$ is 15~nm, so that it is reasonable to assume
that the lubricating layers are composed of only water or a few 
membranes ($\approx 10$), perfectly aligned along the walls. 
In any case, the viscosity
of the film is probably of the order of $10^{-3}$~Pa.s.  

\subsubsection{A possible explanation for the origin of lubricating layers and their
variations with $\sigma$}

\begin{figure}[htbp]
\begin{center}
\scalebox{1}{\includegraphics{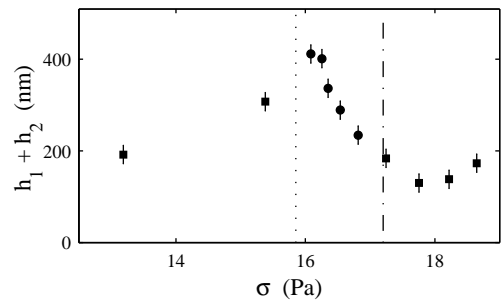}}
\end{center}
\caption{
$h_{\scriptscriptstyle 1}+h_{\scriptscriptstyle 2}$ vs. stress $\sigma$. 
The dotted line indicates $\sigmaA=15.9$~Pa and the dash-dotted line $\sigmaB=17.2$~Pa (see text).
($\blacksquare$) correspond to homogeneous velocity profiles and 
($\bullet$) to banded velocity profiles.
\label{epaisseur_total}}
\end{figure}

In the onion texture, flow can {\it compress} the onions by
changing the smectic period $d$. Such a compression expells some water from
inside the onions and helps to lubricate the flow. This effect has been
shown in a lot of lyotropic lamellar phases using 
neutron and x-ray diffraction \cite{Diat:95,Welch:02}.
In our system, as measured with neutron scattering, the smectic period
changes sligthly over the range of $\gammap$ under study \cite{Diat:95}. If we assume that
the expelled water can also {\it migrate} from the bulk material 
to the walls in order to lubricate the flow, the quantity $h_{\scriptscriptstyle 1}+h_{\scriptscriptstyle 2}$
is then proportional to the volume of water in the lubricating layers.
Figure~\ref{epaisseur_total} presents $h_{\scriptscriptstyle 1}+h_{\scriptscriptstyle 2}$
vs. $\sigma$. Again, the three regions of the 
shear-banding phenomenology can be clearly identified from Fig.~\ref{epaisseur_total}.
At low applied stress, the quantity of water at the walls increases under 
increasing stress. This confirms the assumption that
the viscous stress compresses the onions and expells some water.
At intermediate stress, in the coexistence domain, the volume of water
decreases while the width of the highly sheared band grows in the 
gap of the Couette cell.

Such a result may be simply explained if one considers that some water is needed
to lubricate the flow between the layers of ordered onions. Indeed, as suggested
in Ref.~\cite{Diat:95}, the layering of onions is probably associated with the
expulsion of some water from the onions to lubricate the flow between the layers.
The decrease of $h_{\scriptscriptstyle 1}+h_{\scriptscriptstyle 2}$ could be 
the signature of a {\it migration} of the water from the walls 
to the bulk of the highly sheared band.     
At higher stress, when the layered band has invaded the gap, 
$h_{\scriptscriptstyle 1}+h_{\scriptscriptstyle 2}$ increases again with 
the shear stress $\sigma$. This is consistent with the fact
that the viscous stress tends to compress the onions in the homogeneous layered state.   
Such a picture, suggested by our experiments and by previous works \cite{Diat:95}, is
obviously highly hypothetical. Resolved x-ray or neutron scattering techniques
performed in the gap of the Couette cell are needed to confirm such assumptions.

\section{A simple mechanical approach of the shear-banding instability
\label{asmmotsbi}}

\subsection{Local constitutive flow curve \label{lcfc}}

\begin{figure}[htbp]
\begin{center}
\scalebox{1.0}{\includegraphics{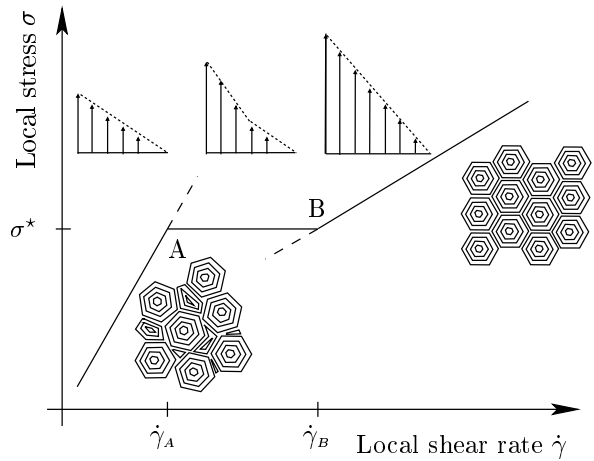}}
\end{center}
\caption{Local constitutive flow curve of the layering transition. 
Two branches corresponding to the two different structures of onions are separated by a 
stress plateau at $\sigma^\star$.
The interface between bands of different shear rates $\gammapA$ and $\gammapB$
is stable at a unique stress $\sigma^\star$. 
The highly sheared band nucleates at point A and invades the gap up to point B
where the flow becomes homogeneous again.}
\label{schematic2}
\end{figure}

The aim of this section is to determine whether the classical mechanical picture
of the shear-banding instability holds in our experiments. In such a picture,
the flow curve is the one sketched in Fig.~\ref{schematic2}. 
Two branches,
corresponding to the two different organizations of onions, are separated by a coexistence
domain. The rheological behaviors of the two branches are given by
$\sigma = f_{i}(\gammap)$, where $i=1$ ($i=2$ resp.) stands for 
the disordered (layered resp.) state of onions. 
In the simple picture of shear-banding, one generally assumes
that the flow field can separate between bands of
different structures, whose local rheological
behaviors are given by $\sigma=f_i(\gammap)$. One also assumes that
the interfaces between the bands are only stable at a critical stress $\sigma^\star$.
When the shear rate is imposed in a geometry where stress is homogeneous
(e.g. a cone-and-plate geometry), such an assumption leads to a {\it stress plateau} on the flow curve at $\sigma^\star$
\cite{Radulescu:00}. 
On the plateau, velocity profiles display
bands of shear rates $\gammapA$ and $\gammapB$. The following equation relates
the imposed {\it global} shear rate to the proportion of bands {\it locally} sheared at
$\gammapA$ and $\gammapB$:
\begin{equation}
\gammap = \alpha\gammapA + (1-\alpha)\gammapB\,, 
\end{equation}
where $\alpha$ is the volume fraction of bands sheared at $\gammapA$. 
In stress-controlled experiments, the plateau cannot be seen anymore and one observes a discontinuous
jump between $\gammapA$ and $\gammapB$ when the stress is sligthly increased above
$\sigma^\star$.
In the next paragraph, we adresss the validity of the assumption that interfaces lie at a critical
stress $\sigma^\star$.

\subsection{Width of the highly sheared band in the Couette geometry}

In the Couette geometry, the stress is not homogeneous inside the gap of the cell and
the simple analysis presented above does not hold anymore.
Indeed, the local stress $\sigma(r)$, where $r$ is the radial position in the 
flow, is given by:
\begin{equation}
\sigma(r) = \frac{\Gamma}{2\pi H r^2} = \sigma_{{\scriptscriptstyle 1}}
\frac{\rot^2}{r^2} \,,
\label{stress_couette}
\end{equation}
where $\Gamma$ is the imposed torque and $\sigma_{{\scriptscriptstyle 1}}=\Gamma/(2 \pi H \rot^2)$ is the
stress at the rotor. 
The shear-banding scenario then leads to a picture with only two bands: one sheared at $\gammapA$ and
one sheared at $\gammapB$.
If one considers that the interface between bands is only stable
at a given stress $\sigma^\star$, one easily calculates the width $\delta$ of the
highly sheared band: 
\begin{equation}
\delta = \rot\left(\sqrt{\frac{\sigma_{\scriptscriptstyle 1}}{\sigma^\star}}-1\right)\,.
\label{hbande_couette}
\end{equation}
Such a relation shows that the stress should increase slightly along   
the coexistence domain. Indeed, when entering the coexistence region, 
i.e. when $\delta=0$, one finds $\sigma_{\scriptscriptstyle 1}=\sigma^\star$,
and when the highly sheared band has invaded the gap, $\delta=e$ and
$\sigma_{\scriptscriptstyle 1} = (\sta/\rot)^2 \sigma^\star$.
This means that, in the Couette geometry, the plateau is not perfectly flat and presents a slope 
$(\sta/\rot)^2 - 1 \approx 2e/\rot$ due to the stress inhomogeneity \cite{Radulescu:00}.

\begin{figure}[htbp]
\begin{center}
\scalebox{1}{\includegraphics{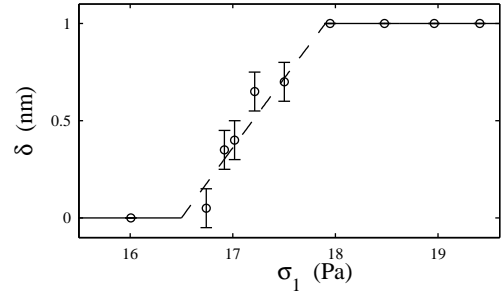}}
\end{center}
\caption{
Width $\delta$ of the highly sheared band vs. $\sigma_{\scriptscriptstyle 1}$ the stress at the rotor.
The dashed line corresponds to Eq.~(\ref{hbande_couette}) with $\sigma^\star = 16.5$~Pa.
\label{bande_T30}}
\end{figure}

Our velocimetry measurements easily yield the width $\delta$ of the highly sheared band. 
Figure~\ref{bande_T30} displays the measured $\delta$ vs. the measured
stationary stress at the rotor $\sigma_{\scriptscriptstyle 1}$. 
We have also added on this figure the theoretical relation of Eq.~(\ref{hbande_couette})
with $\sigma^\star=16.5$~Pa, which yields
a satisfactory agreement with the experimental data. 
This seems to indicate that the stability criterion for the interface 
holds in our experiments with $\sigma^\star=16.5$~Pa.
However, our data presents relatively large errorbars on the measured $\delta$.
To proceed further in the analysis and to confirm the mechanical picture of the 
shear-banding instability, one should now try to
fit all the measured velocity profiles from the knowledge of the flow curve.  

\subsection{Procedure for fitting the velocity profiles}

Our fitting procedure relies on the following relations:
\begin{align}
\sigma(r) &= \frac{\Gamma}{2\pi H r^2} \,, \label{chang_var1} \\
\gammap(r) &= r\,\frac{\partial}{\partial r} \left(\frac{v}{r}\right)\,.
\label{chang_var2}
\end{align}  
In order to compute velocity profiles from Eqs.~(\ref{chang_var1}) and (\ref{chang_var2}), 
we first have to determine the rheological laws linking the local
stress $\sigma(r)$ and the local shear rate $\gammap(r)$.

\subsubsection{Rheological behaviors of the homogeneous branches \label{sss:rbothb}}

\begin{figure}[htbp]
\begin{center}
\scalebox{1}{\includegraphics{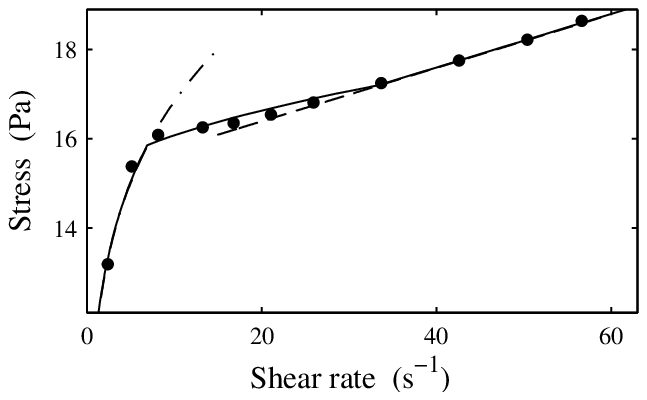}}
\end{center}
\caption{
($\bullet$) Effective flow curve $\sigma$ vs. $\gammap_{\hbox{{\tiny eff}}}$ 
displayed in Fig.~\ref{t30rheol}.
The dash-dotted line is the best fit of the 
low sheared branch ($\gammap \leq6$~s$^{-1}$) 
using a shear-thinning behavior $\sigma=A\gammap^n$ with $A=11.49$ and $n=0.17$.
The dashed line is the best fit of the highly sheared branch ($\gammap\geq30$~s$^{-1}$) 
according to a Bingham behavior $\sigma = \seuil +\eta\gammap$ 
with $\seuil=15.19$~Pa and $\eta=0.06$~Pa.s. 
The continuous line is computed from Eqs.~(\ref{int1})--(\ref{int3}) 
with $\sigma^\star = 16.5$~Pa.
\label{Rheol_correc_T30}}
\end{figure}

Figure~\ref{Rheol_correc_T30} presents the effective flow curve $\sigma$($\gammapeff$) measured in our
experiment at $T=30^\circ$C and under controlled shear rate.
We have fitted the two branches corresponding to the two homogeneous states:
(i) for $\gammap \leq6$~s$^{-1}$ (disordered onions) by a 
shear-thinning behavior 
$\sigma= f_{\scriptscriptstyle 1}(\gammap) = A\gammap^n$ with $A=11.49$ and $n=0.17$;
(ii) and for $\gammap\geq30$~s$^{-1}$ (layered state) by
a Bingham fluid $\sigma = f_{\scriptscriptstyle 2}(\gammap) = \seuil +\eta\gammap$ 
with $\seuil=15.19$~Pa and $\eta=0.06$~Pa.s. The resulting fits are 
presented in Fig.~\ref{Rheol_correc_T30}.  
Since the flow field is homogeneous on these branches, such behaviors
fitted from the effective flow curve, i.e. after contributions due to wall
slip are removed, represent
the local rheological behavior $\sigma$ vs. $\gammap$.

\subsubsection{Equations for the theoretical flow curve and the velocity profiles}

Let us now assume that the interface
between the bands lies at a unique stress $\sigma^\star$ and let us consider a given torque $\Gamma$.
From $\Gamma$ and Eq.~(\ref{chang_var1}), we get the local stress $\sigma(r)$.
If $\sigma(r)<\sigma^\star$ everywhere in the gap of the Couette cell, the flow is homogeneous
and composed of the disordered onions, whereas
if $\sigma(r)>\sigma^\star$ everywhere, the flow
is also homogeneous but composed of the layered state of onions. 
Moreover, the local shear rate  $\gammap(r)$ is given by Eq.~(\ref{chang_var2}), so that in both cases, 
the velocity profiles are given by the following integration of the rheological
behaviors:
\begin{equation}
\frac{v(r)}{r} = \frac{v_{\scriptscriptstyle 2}}{\sta}+\int_{\sta}^{r}
\frac{\gammap(u)}{u}\,\hbox{d}u\,,
\label{int1}
\end{equation}     
where $\gammap(r)$ is found by solving:
\begin{align}
\sigma(r)= \frac{\Gamma}{2\pi H r^2}=f_i\left(\gammap(r)\right)\,, 
\end{align}
and $i$ denotes the considered branch.

When there exists one particular position in the gap where $\sigma(r)=\sigma^\star$,
the flow is inhomogeneous and composed of two different bands. To calculate the resulting velocity profile,
one should separate the previous integration according to:
\begin{align}
\frac{v(r)}{r} &=  \frac{v_{\scriptscriptstyle 2}}{\sta}+\int_{\sta}^{r}\frac{\gammap(u)}{u} \hbox{d}u\,\,\,  
\hbox{for\, $r>\rot+\delta$ and} \notag \\[4mm]
\frac{v(r)}{r} &= \frac{v(\rot+\delta)}{\rot+\delta}+ \int_{\rot+\delta}^{r}\frac{\gammap(u)}{u} \hbox{d}u\,\,\, 
\hbox{for\, $r<\rot+\delta$} \,,
\label{int2}
\end{align}
where $\delta$ is given by Eq.~(\ref{hbande_couette}), and $\gammap(r)$ by:
\begin{align}
\sigma(r)&=f_{\scriptscriptstyle 1}\left(\gammap(r)\right)\,\, \hbox{for $r>\rot+\delta$,} \notag \\[3mm]
\sigma(r)&=f_{\scriptscriptstyle 2}\left(\gammap(r)\right)\,\, \hbox{for $r<\rot+\delta$} \,.
\label{int3}
\end{align}
The Appendix presents the detailed integration of Eqs.~(\ref{int1})--(\ref{int3})
for the specific behaviors $\sigma=f_{\scriptscriptstyle 1}(\gammap)$ and $\sigma=f_{\scriptscriptstyle 2}(\gammap)$
obtained previously.
Note that Eqs.~(\ref{int1})--(\ref{int3}) require the knowledge of $v_{s\scriptscriptstyle 2}=v_{\scriptscriptstyle 2}$ 
for the arbitrary torque $\Gamma$. Thus, to compute $v_{s\scriptscriptstyle 2}$ at any value of $\Gamma$, we first 
interpolate the data $v_{si}$ vs. $\sigma$ displayed in Fig.~\ref{vs_sigma_T30}(a).  

For each value of $\Gamma$, i.e. for each value of $\sigma$ indicated by the rheometer 
[see Eq.~(\ref{e.sigmarheo})], one can calculate a theoretical velocity profile
$v(x)$ using the fitting procedure detailed above. 
One can then calculate the corresponding effective shear rate $\gammapeff$ using Eq.~(\ref{e.effectiveshear}).    
Such a procedure allows us to compute a theoretical flow curve $\sigma$ vs. $\gammapeff$. 
This theoretical flow curve is displayed in Fig.~\ref{Rheol_correc_T30} for 
comparison with the measured rheological data.
The coexistence domain is well reproduced by this theoretical effective flow curve, namely the slope of the 
stress plateau. This good agreement confirms the assumption that the interface between the bands
lies at a given value of the stress $\sigma^\star$. Note that the value $\sigma^\star=16.5$~Pa is the only
free parameter of the fitting procedure.
The value of $\sigma^\star$ also yields the value of the stress $\sigmaA$ ($\sigmaB$ resp.) at the entrance 
(at the end resp.) of the coexistence domain:
\begin{align}
\sigmaA &= \frac{\rot^2+\sta^2}{2\sta^2}\, \sigma^\star \label{siga_sigb1} \approx 15.9~\hbox{Pa} \,,\\[3mm] 
\sigmaB &= \frac{\rot^2+\sta^2}{2\rot^2}\, \sigma^\star \approx 17.2~\hbox{Pa} \,. 
\label{siga_sigb2}
\end{align}    
The above values of $\sigmaA$ and $\sigmaB$ are those displayed in Fig.~\ref{vs_sigma_T30}, 
\ref{hs_sigma_T30} and \ref{epaisseur_total} 
to visualize the coexistence domain.

\begin{figure}[htbp]
\begin{center}
\scalebox{1}{\includegraphics{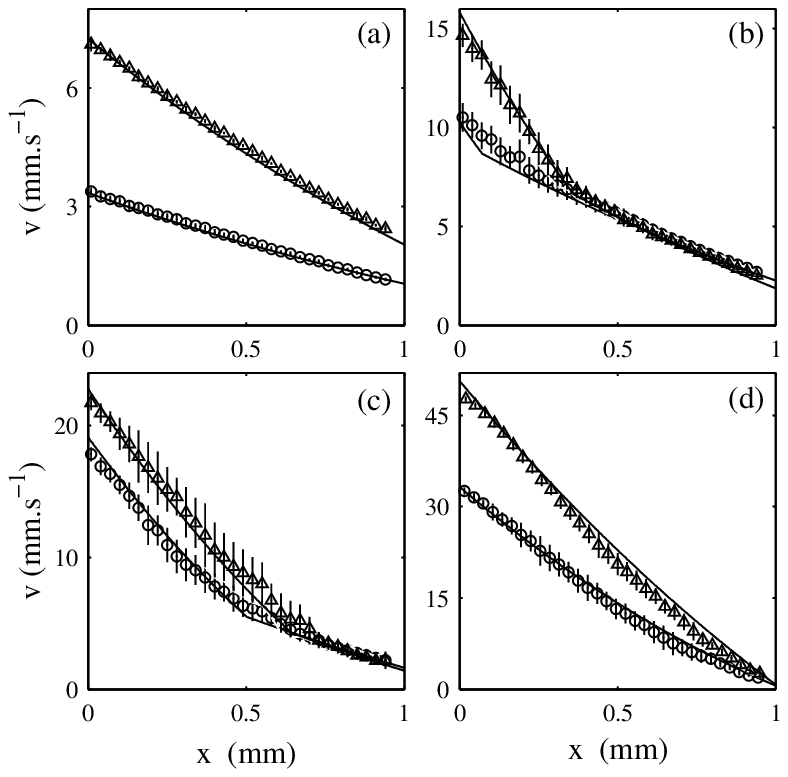}}
\end{center}
\caption{Time-averaged velocity profiles displayed in Fig.~\ref{profilt30}.
$\gammap=5$ ($\circ$) and
$10$~s$^{-1}$ ($\triangle$).
(b) 
$\gammap=15$ ($\circ$), and
$20$~s$^{-1}$ ($\triangle$).
(c) 
$\gammap=22.5$ ($\circ$) and 
$26$~s$^{-1}$ ($\triangle$).
(d)
$\gammap=37$ ($\circ$) and
$53$~s$^{-1}$ ($\triangle$).
The errorbars represent the temporal fluctuations of the local velocity. 
The continous lines are the theoretical velocity profiles calculated from Eqs.~(\ref{int1})--(\ref{int3}).
\label{fit_T30}}
\end{figure}

For each value of $\Gamma$ and thus for each theoretical velocity profile, one can also calculate 
the shear rate $\gammap$ indicated by the rheometer. Indeed, it is easy to determine the rotor velocity
from $v(x)$ and from the values of $v_{s\scriptscriptstyle 1}$ displayed in Fig.~\ref{vs_sigma_T30}(b):
$v_{\scriptscriptstyle 0} = v(x\!=\!0)+v_{s\scriptscriptstyle 1}$. 
$\gammap$ is then given by Eq.~(\ref{e.gammarheo}).  
From the calculated velocity profiles, we extracted those corresponding to the shear rates applied
in our experiments. Figure~\ref{fit_T30} shows these theoretical fits $v(x)$ and the 
corresponding experimental velocity profiles. Again, the good agreement indicates that the  
stability criterion for the interface holds in our experiments. Such an agreement also confirms the
observed shear-banding phenomenology and more precisely, the fact that near the layering transition, 
the flow is composed of two different bands characterized by two different rheological behaviors.     

\subsection{Discussion on the validity of the mechanical approach \label{ss:vataisrais}} 
In the previous mechanical approach, it is important to note that to each value of torque $\Gamma$ and
thus to each value of stress $\sigma$, we can associate a unique velocity profile $v(x)$, and thus a 
unique shear rate $\gammap$. This is a direct consequence of the slope of the stress plateau of the flow curve
in the Couette geometry. The uniqueness of $\gammap$ for a given $\sigma$ means that 
the shear-banding phenomenology is also expected when the stress is controlled in a Couette geometry.
However, as mentionned in Sec.~\ref{ssec:tlt:ssarc}
(see Fig.~\ref{schematic1}), the responses of the shear rate under controlled 
stress for temperatures $T\geq T_c=27^\circ$C, are no
longer stationary and present large oscillations in the vicinity of the layering transition.
It is thus obvious that some ingredients are missing in the mechanical approach detailed previously to model such dynamics.  
For temperatures $T<T_c$, 
the responses of the shear rate under controlled stress are almost stationary and only display
relatively small fluctuations at the layering transition. 
Our mechanical approach may thus be fully validated only at $T<T_c$. This led us to repeat our experiments
at $T=26^\circ$C both under controlled stress and under controlled shear rate.

\begin{figure}[htbp]
\begin{center}
\scalebox{1}{\includegraphics{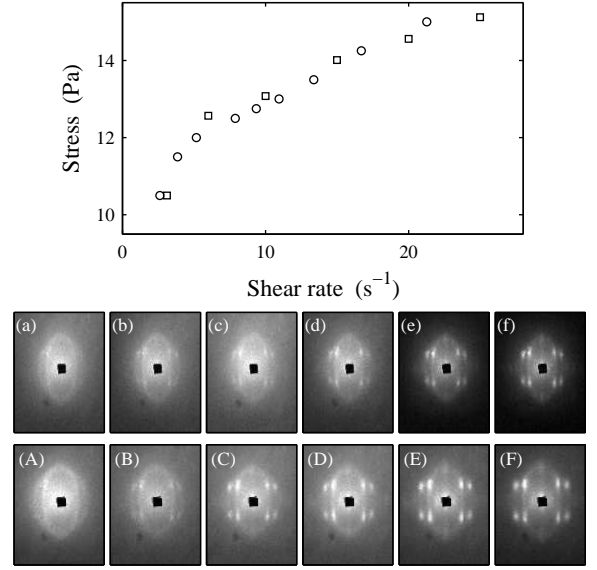}}
\end{center}
\caption{Stationary flow curves obtained at $T=26^\circ$C under controlled stress ($\circ$)
and under controlled shear rate ($\square$).
Corresponding diffraction patterns at imposed stress:
(a) $\sigma=12$, 
(b) 12.5,
(c) 12.75,
(d) 13,
(e) 13.5, and
(f) 14.25~Pa.
Corresponding diffraction patterns at imposed shear rate:
(A) $\gammap=6$,
(B) 10,
(C) 15,
(D) 20,
(E) 25, and
(F) 30~s$^{-1}$.\label{rheol_T26}}
\end{figure}

\begin{figure}[htbp]
\begin{center}
\scalebox{1}{\includegraphics{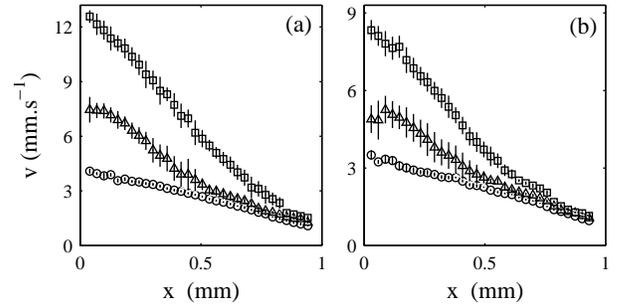}}
\end{center}
\caption{Time-averaged velocity profiles measured simultaneously to the flow curves displayed in Fig.~\ref{rheol_T26}
at $T=26^\circ$C.
The errorbars represent the temporal fluctuations of the local velocity.  
(a) Imposed shear rate:
$\gammap=6$ ($\circ$),
$10$ ($\triangle$), and
$15$~s$^{-1}$ ($\square$).
(b) Imposed stress: 
$\sigma=12$ ($\circ$),
12.5 ($\triangle$), and
13~Pa ($\square$).
\label{prof_T26}}
\end{figure}

Using careful protocols similar to that of Sec.~\ref{gralv}, we measure the two 
stationary flow curves displayed in Fig.~\ref{rheol_T26}, as well as the corresponding diffraction patterns.  
Figure~\ref{prof_T26} presents the time-averaged velocity profiles measured simultaneously to the flow curves
displayed in Fig.~\ref{rheol_T26}.  
These experiments clearly demonstrate that the phenomenology of shear-banding, previously observed           
at $T=30^\circ$C under controlled $\gammap$, is also recovered at $T=26^\circ$C both at imposed $\gammap$ and 
imposed $\sigma$. 
As expected from the mechanical approach, there is no difference 
between controlling the stress or the shear rate in the Couette geometry, as long as 
the rheological responses are stationary. 

\section{Conclusions \label{conclusion}}

In this paper, we have presented an extensive study of a shear-induced transition in a lyotropic system:
the layering transition of the onion texture. Using  velocimetry, 
structural measurements and rheological data, we have shown that 
the classical picture of {\it shear-banding} holds for the layering transition. In this
picture, velocity profiles display bands in the vicinity of the transition and the nucleation 
and growth of the highly sheared band is associated to the nucleation and growth of the SIS, 
i.e. of the layered state of onions. Using the classical mechanical approach \cite{Spenley:93,Lu:00}, 
in which the interface 
between the bands lies at a given value of the stress, we were
able to fit the velocity profiles and the coexistence domain on the flow curve. 

Moreover, our data reveal the presence of strong wall slip. We have shown that the variations of wall slip with 
stress are in good agreement with the shear-banding behavior: when the flow is homogeneous, the slip
velocity is a universal function of $\sigma$ since inertial effects are negligible, 
and when the flow displays bands, wall slip presents
a very large dissymmetry at the rotor and at the stator, probably because the two different textures have different
behaviors at the walls of the Couette cell.

One of the most puzzling features of our results is the presence of large temporal fluctuations of the flow field in the coexistence
domain. Indeed, velocity profiles display localized temporal fluctuations reaching up to 20\% at the interface between
the two bands. Moreover, the characteristic times of such fluctuations are very long (100--1000~s), and 
do not correspond to external mechanical vibrations of the Couette cell. 
In the next part of the paper \cite{Salmon:03_5}, we
analyze these temporal fluctuations and show that they are probably the signature 
of {\it rheochaos}.            

\appendix
\section{}
In the particular case of the rheogical behaviors obtained in Sec.~\ref{sss:rbothb}, i.e.
the disordered texture $\sigma=f_{\scriptscriptstyle 1}(\gammap)$ has a shear-thinning behavior $\sigma=A\gammap^n$ and
the layered state $\sigma=f_{\scriptscriptstyle 2}(\gammap)$ 
is a Bingham fluid $\sigma=\seuil+\eta\gammap$, Eqs.~(\ref{int1})--(\ref{int3}) read:\\
\noindent
(i) if $\sigma(r) < \sigma^\star$, the velocity profile is given by:  
\begin{align} 
v(x) = v_{s{\scriptscriptstyle 2}} \frac{r}{\sta} +r \frac{n}{2} 
\left[ \frac{\Gamma}{2\pi H \sta^2\,A} \right]^{1/n} 
\left[ \left(\frac{R_{\scriptscriptstyle 2}}{r}\right)^{2/n} -1\right];
\notag 
\end{align}
(ii) if $\sigma(r) > \sigma^\star$, the flow is also homogeneous and the velocity profile reads:   
\begin{align} 
&v(x) = v_{s{\scriptscriptstyle 2}} \frac{r}{\sta}+r\,\frac{\Gamma}{4\pi\,H\,\sta^2\,\eta}
\left[ \left(\frac{R_{\scriptscriptstyle 2}}{r}\right)^{2} -1\right] +  \notag \\ 
&r\,\frac{\seuil}{\eta}\,
\ln\left(\frac{r}{\sta}\right) \,;  
\notag
\end{align}
(iii) if there exists $r$ for which $\sigma(r) = \sigma^\star$, the flow displays two bands, and for $\rot+\delta<r<\sta$:
\begin{align}
v(x) &= v_{s{\scriptscriptstyle 2}} \frac{r}{\sta}+r \frac{n}{2} 
\left[ \frac{\Gamma}{2\pi H \sta^2\,A} \right]^{1/n} 
\left[ \left(\frac{\sta}{r}\right)^{2/n} -1\right], \notag \\ \notag
\notag
\end{align}
whereas for $\rot<r<\rot+\delta$:
\begin{align}
&v(x) = \,\,r\,\frac{\Gamma}{4\pi\,H\,(\rot+\delta)^2\,\eta} \left[ 
\left(\frac{\rot+\delta}{r}\right)^{2} -1\right]+ \notag \\  
& r\,\frac{\seuil}{\eta}\,\ln\left(\frac{r}{\rot+\delta}\right) + \notag \\  
& v_{s{\scriptscriptstyle 2}}\,\frac{r}{\sta}+r\,\frac{n}{2} 
\left[ \frac{\Gamma}{2\pi\,H\,\sta^2\,A} \right]^{1/n} 
\left[ \left(\frac{\sta}{\rot+\delta}\right)^{2/n} -1\right]\,. \notag
\end{align}

\begin{acknowledgments}
The authors are deeply grateful to D. Roux, 
L. B\'ecu, and C. Gay for many fruitful discussions and comments on this work. 
We also thank Pr. F. Nadus for helping us with the figures of this paper. 
The authors would like to thank the {\it Cellule Instrumentation} at CRPP for the realization of the heterodyne DLS setup.
\end{acknowledgments}

\bibliographystyle{apsrev}

\begin{thebibliography}{47}
\expandafter\ifx\csname natexlab\endcsname\relax\def\natexlab#1{#1}\fi
\expandafter\ifx\csname bibnamefont\endcsname\relax
  \def\bibnamefont#1{#1}\fi
\expandafter\ifx\csname bibfnamefont\endcsname\relax
  \def\bibfnamefont#1{#1}\fi
\expandafter\ifx\csname citenamefont\endcsname\relax
  \def\citenamefont#1{#1}\fi
\expandafter\ifx\csname url\endcsname\relax
  \def\url#1{\texttt{#1}}\fi
\expandafter\ifx\csname urlprefix\endcsname\relax\def\urlprefix{URL }\fi
\providecommand{\bibinfo}[2]{#2}
\providecommand{\eprint}[2][]{\url{#2}}

\bibitem[{\citenamefont{Larson}(1999)}]{Larson:99}
\bibinfo{author}{\bibfnamefont{R.~G.} \bibnamefont{Larson}},
  \emph{\bibinfo{title}{The Structure and Rheology of Complex Fluids}}
  (\bibinfo{publisher}{Oxford University Press}, \bibinfo{year}{1999}).

\bibitem[{\citenamefont{Guyon et~al.}(1994)\citenamefont{Guyon, Hulin, and
  Petit}}]{Guyon:94}
\bibinfo{author}{\bibfnamefont{E.}~\bibnamefont{Guyon}},
  \bibinfo{author}{\bibfnamefont{J.-P.} \bibnamefont{Hulin}}, \bibnamefont{and}
  \bibinfo{author}{\bibfnamefont{L.}~\bibnamefont{Petit}},
  \emph{\bibinfo{title}{Hydrodynamique physique}}
  (\bibinfo{publisher}{Inter\'Editions/\'Editions du CNRS},
  \bibinfo{year}{1994}).

\bibitem[{\citenamefont{Cates and Evans}(2000)}]{Edimbourg:00}
\bibinfo{editor}{\bibfnamefont{M.~E.} \bibnamefont{Cates}} \bibnamefont{and}
  \bibinfo{editor}{\bibfnamefont{M.~R.} \bibnamefont{Evans}}, eds.,
  \emph{\bibinfo{title}{Soft and Fragile Matter: Non Equilibrium Dynamics
  Metastability and Flow}} (\bibinfo{publisher}{Institute of Physics Publishing
  (Bristol)}, \bibinfo{year}{2000}).

\bibitem[{\citenamefont{H\'{e}braud et~al.}(1997)\citenamefont{H\'{e}braud,
  Lequeux, Munch, and Pine}}]{Hebraud:97}
\bibinfo{author}{\bibfnamefont{P.}~\bibnamefont{H\'{e}braud}},
  \bibinfo{author}{\bibfnamefont{F.}~\bibnamefont{Lequeux}},
  \bibinfo{author}{\bibfnamefont{J.~P.} \bibnamefont{Munch}}, \bibnamefont{and}
  \bibinfo{author}{\bibfnamefont{D.~J.} \bibnamefont{Pine}},
  \bibinfo{journal}{Phys. Rev. Lett.} \textbf{\bibinfo{volume}{78}},
  \bibinfo{pages}{4657} (\bibinfo{year}{1997}).

\bibitem[{\citenamefont{Berret et~al.}(1994)\citenamefont{Berret, Roux, and
  Porte}}]{Berret:94}
\bibinfo{author}{\bibfnamefont{J.-F.} \bibnamefont{Berret}},
  \bibinfo{author}{\bibfnamefont{D.~C.} \bibnamefont{Roux}}, \bibnamefont{and}
  \bibinfo{author}{\bibfnamefont{G.}~\bibnamefont{Porte}}, \bibinfo{journal}{J.
  Phys. II} \textbf{\bibinfo{volume}{4}}, \bibinfo{pages}{1261}
  (\bibinfo{year}{1994}).

\bibitem[{\citenamefont{Schmitt et~al.}(1994)\citenamefont{Schmitt, Lequeux,
  Pousse, and Roux}}]{Schmitt:94}
\bibinfo{author}{\bibfnamefont{V.}~\bibnamefont{Schmitt}},
  \bibinfo{author}{\bibfnamefont{F.}~\bibnamefont{Lequeux}},
  \bibinfo{author}{\bibfnamefont{A.}~\bibnamefont{Pousse}}, \bibnamefont{and}
  \bibinfo{author}{\bibfnamefont{D.}~\bibnamefont{Roux}},
  \bibinfo{journal}{Langmuir} \textbf{\bibinfo{volume}{10}},
  \bibinfo{pages}{955} (\bibinfo{year}{1994}).

\bibitem[{\citenamefont{Diat et~al.}(1993)\citenamefont{Diat, Roux, and
  Nallet}}]{Diat:93_2}
\bibinfo{author}{\bibfnamefont{O.}~\bibnamefont{Diat}},
  \bibinfo{author}{\bibfnamefont{D.}~\bibnamefont{Roux}}, \bibnamefont{and}
  \bibinfo{author}{\bibfnamefont{F.}~\bibnamefont{Nallet}},
  \bibinfo{journal}{J. Phys. II (France)} \textbf{\bibinfo{volume}{3}},
  \bibinfo{pages}{1427} (\bibinfo{year}{1993}).

\bibitem[{\citenamefont{Roux et~al.}(1993)\citenamefont{Roux, Nallet, and
  Diat}}]{Roux:93}
\bibinfo{author}{\bibfnamefont{D.}~\bibnamefont{Roux}},
  \bibinfo{author}{\bibfnamefont{F.}~\bibnamefont{Nallet}}, \bibnamefont{and}
  \bibinfo{author}{\bibfnamefont{O.}~\bibnamefont{Diat}},
  \bibinfo{journal}{Europhys. Lett.} \textbf{\bibinfo{volume}{24}},
  \bibinfo{pages}{53} (\bibinfo{year}{1993}).

\bibitem[{\citenamefont{Hu et~al.}(1998)\citenamefont{Hu, Boltenhagen, and
  Pine}}]{Hu_1:98}
\bibinfo{author}{\bibfnamefont{Y.~T.} \bibnamefont{Hu}},
  \bibinfo{author}{\bibfnamefont{P.}~\bibnamefont{Boltenhagen}},
  \bibnamefont{and} \bibinfo{author}{\bibfnamefont{D.~J.} \bibnamefont{Pine}},
  \bibinfo{journal}{J. Rheol.} \textbf{\bibinfo{volume}{42}},
  \bibinfo{pages}{1185} (\bibinfo{year}{1998}).

\bibitem[{\citenamefont{Eiser et~al.}(2000)\citenamefont{Eiser, Molino,
  Porte, and Diat}}]{Eiser:00}
\bibinfo{author}{\bibfnamefont{E.}~\bibnamefont{Eiser}},
  \bibinfo{author}{\bibfnamefont{F.}~\bibnamefont{Molino}}, 
  \bibinfo{author}{\bibfnamefont{G.}~\bibnamefont{Porte}}, \bibnamefont{and}
  \bibinfo{author}{\bibfnamefont{O.}~\bibnamefont{Diat}},
\bibinfo{journal}{Phys. Rev. E} \textbf{\bibinfo{volume}{61}},
  \bibinfo{pages}{6759} (\bibinfo{year}{2000}).

\bibitem[{\citenamefont{Ramos et~al.}(2000)\citenamefont{Ramos, Molino, and
  Porte}}]{Ramos:00}
\bibinfo{author}{\bibfnamefont{L.}~\bibnamefont{Ramos}},
  \bibinfo{author}{\bibfnamefont{F.}~\bibnamefont{Molino}}, \bibnamefont{and}
  \bibinfo{author}{\bibfnamefont{G.}~\bibnamefont{Porte}},
  \bibinfo{journal}{Langmuir} \textbf{\bibinfo{volume}{16}},
  \bibinfo{pages}{5846} (\bibinfo{year}{2000}).

\bibitem[{\citenamefont{Spenley et~al.}(1993)\citenamefont{Spenley, Cates, and
  McLeish}}]{Spenley:93}
\bibinfo{author}{\bibfnamefont{N.~A.} \bibnamefont{Spenley}},
  \bibinfo{author}{\bibfnamefont{M.~E.} \bibnamefont{Cates}}, \bibnamefont{and}
  \bibinfo{author}{\bibfnamefont{T.~C.~B.} \bibnamefont{McLeish}},
  \bibinfo{journal}{Phys. Rev. Lett.} \textbf{\bibinfo{volume}{71}},
  \bibinfo{pages}{939} (\bibinfo{year}{1993}).

\bibitem[{\citenamefont{Kabla and Debr\'egeas}(2003)}]{Kabla:03}
\bibinfo{author}{\bibfnamefont{A.}~\bibnamefont{Kabla}} \bibnamefont{and}
  \bibinfo{author}{\bibfnamefont{G.}~\bibnamefont{Debr\'egeas}},
  \bibinfo{journal}{Phys. Rev. Lett.} \textbf{\bibinfo{volume}{90}},
  \bibinfo{pages}{258303} (\bibinfo{year}{2003}).

\bibitem[{\citenamefont{Olmsted and Goldbart}(1992)}]{Olmsted:92}
\bibinfo{author}{\bibfnamefont{P.~D.} \bibnamefont{Olmsted}} \bibnamefont{and}
  \bibinfo{author}{\bibfnamefont{P.~M.} \bibnamefont{Goldbart}},
  \bibinfo{journal}{Phys. Rev. A} \textbf{\bibinfo{volume}{46}},
  \bibinfo{pages}{4966} (\bibinfo{year}{1992}).

\bibitem[{\citenamefont{Lu et~al.}(2000)\citenamefont{Lu, Olmsted, and
  Ball}}]{Lu:00}
\bibinfo{author}{\bibfnamefont{C.-Y.~D.} \bibnamefont{Lu}},
  \bibinfo{author}{\bibfnamefont{P.~D.} \bibnamefont{Olmsted}},
  \bibnamefont{and} \bibinfo{author}{\bibfnamefont{R.~C.} \bibnamefont{Ball}},
  \bibinfo{journal}{Phys. Rev. Lett.} \textbf{\bibinfo{volume}{84}},
  \bibinfo{pages}{642} (\bibinfo{year}{2000}).

\bibitem[{\citenamefont{Goveas and Olmsted}(2001)}]{Goveas:01}
\bibinfo{author}{\bibfnamefont{J.~L.} \bibnamefont{Goveas}} \bibnamefont{and}
  \bibinfo{author}{\bibfnamefont{P.~D.} \bibnamefont{Olmsted}},
  \bibinfo{journal}{Eur. Phys. J. E} \textbf{\bibinfo{volume}{6}},
  \bibinfo{pages}{79} (\bibinfo{year}{2001}).

\bibitem[{\citenamefont{Yuan}(1999)}]{Yuan:99}
\bibinfo{author}{\bibfnamefont{X.-F.} \bibnamefont{Yuan}},
  \bibinfo{journal}{Europhys. Lett.} \textbf{\bibinfo{volume}{46}},
  \bibinfo{pages}{542} (\bibinfo{year}{1999}).

\bibitem[{\citenamefont{Ajdari}(1998)}]{Ajdari:98}
\bibinfo{author}{\bibfnamefont{A.}~\bibnamefont{Ajdari}},
  \bibinfo{journal}{Phys. Rev. E} \textbf{\bibinfo{volume}{58}},
  \bibinfo{pages}{6294} (\bibinfo{year}{1998}).

\bibitem[{\citenamefont{Picard et~al.}(2002)\citenamefont{Picard, Ajdari,
  Bocquet, and Lequeux}}]{Picard:02}
\bibinfo{author}{\bibfnamefont{G.}~\bibnamefont{Picard}},
  \bibinfo{author}{\bibfnamefont{A.}~\bibnamefont{Ajdari}},
  \bibinfo{author}{\bibfnamefont{L.}~\bibnamefont{Bocquet}}, \bibnamefont{and}
  \bibinfo{author}{\bibfnamefont{F.}~\bibnamefont{Lequeux}},
  \bibinfo{journal}{Phys. Rev. E} \textbf{\bibinfo{volume}{66}},
  \bibinfo{pages}{051501} (\bibinfo{year}{2002}).

\bibitem[{\citenamefont{Cappelaere et~al.}(1997)\citenamefont{Cappelaere,
  Berret, Decruppe, Cressely, and Lindner}}]{Cappelaere:97}
\bibinfo{author}{\bibfnamefont{E.}~\bibnamefont{Cappelaere}},
  \bibinfo{author}{\bibfnamefont{J.-F.} \bibnamefont{Berret}},
  \bibinfo{author}{\bibfnamefont{J.-P.} \bibnamefont{Decruppe}},
  \bibinfo{author}{\bibfnamefont{R.}~\bibnamefont{Cressely}}, \bibnamefont{and}
  \bibinfo{author}{\bibfnamefont{P.}~\bibnamefont{Lindner}},
  \bibinfo{journal}{Phys. Rev. E} \textbf{\bibinfo{volume}{56}},
  \bibinfo{pages}{1869} (\bibinfo{year}{1997}).

\bibitem[{\citenamefont{Lerouge et~al.}(1998)\citenamefont{Lerouge, Decruppe,
  and Humbert}}]{Lerouge:98}
\bibinfo{author}{\bibfnamefont{S.}~\bibnamefont{Lerouge}},
  \bibinfo{author}{\bibfnamefont{J.-P.} \bibnamefont{Decruppe}},
  \bibnamefont{and} \bibinfo{author}{\bibfnamefont{C.}~\bibnamefont{Humbert}},
  \bibinfo{journal}{Phys. Rev. Lett.} \textbf{\bibinfo{volume}{81}},
  \bibinfo{pages}{5457} (\bibinfo{year}{1998}).

\bibitem[{\citenamefont{Mair and Callaghan}(1996)}]{Mair:96}
\bibinfo{author}{\bibfnamefont{R.~W.} \bibnamefont{Mair}} \bibnamefont{and}
  \bibinfo{author}{\bibfnamefont{P.~T.} \bibnamefont{Callaghan}},
  \bibinfo{journal}{Europhys. Lett.} \textbf{\bibinfo{volume}{36}},
  \bibinfo{pages}{719} (\bibinfo{year}{1996}).

\bibitem[{\citenamefont{Britton and Callaghan}(1999)}]{Britton:99}
\bibinfo{author}{\bibfnamefont{M.~M.} \bibnamefont{Britton}} \bibnamefont{and}
  \bibinfo{author}{\bibfnamefont{P.~T.} \bibnamefont{Callaghan}},
  \bibinfo{journal}{Eur. Phys. J. B} \textbf{\bibinfo{volume}{7}},
  \bibinfo{pages}{237} (\bibinfo{year}{1999}).

\bibitem[{\citenamefont{Fischer and Callaghan}(2001)}]{Fischer:01}
\bibinfo{author}{\bibfnamefont{E.}~\bibnamefont{Fischer}} \bibnamefont{and}
  \bibinfo{author}{\bibfnamefont{P.~T.} \bibnamefont{Callaghan}},
  \bibinfo{journal}{Phys. Rev. E} \textbf{\bibinfo{volume}{64}},
  \bibinfo{pages}{011501} (\bibinfo{year}{2001}).

\bibitem[{\citenamefont{Salmon et~al.}(2003{\natexlab{a}})\citenamefont{Salmon,
  Colin, Manneville, and Molino}}]{Salmon:03_4}
\bibinfo{author}{\bibfnamefont{J.-B.} \bibnamefont{Salmon}},
 \bibinfo{author}{\bibfnamefont{A.}~\bibnamefont{Colin}}, 
  \bibinfo{author}{\bibfnamefont{S.}~\bibnamefont{Manneville}}, \bibnamefont{and}
  \bibinfo{author}{\bibfnamefont{F.}~\bibnamefont{Molino}},
  \bibinfo{journal}{Phys. Rev. Lett.} \textbf{\bibinfo{volume}{90}},
  \bibinfo{pages}{228303} (\bibinfo{year}{2003}{\natexlab{a}}).

\bibitem[{\citenamefont{Diat et~al.}(1995)\citenamefont{Diat, Roux, and
  Nallet}}]{Diat:95}
\bibinfo{author}{\bibfnamefont{O.}~\bibnamefont{Diat}},
  \bibinfo{author}{\bibfnamefont{D.}~\bibnamefont{Roux}}, \bibnamefont{and}
  \bibinfo{author}{\bibfnamefont{F.}~\bibnamefont{Nallet}},
  \bibinfo{journal}{Phys. Rev. E} \textbf{\bibinfo{volume}{51}},
  \bibinfo{pages}{3296} (\bibinfo{year}{1995}).

\bibitem[{\citenamefont{Sierro and Roux}(1997)}]{Sierro:97}
\bibinfo{author}{\bibfnamefont{P.}~\bibnamefont{Sierro}} \bibnamefont{and}
  \bibinfo{author}{\bibfnamefont{D.}~\bibnamefont{Roux}},
  \bibinfo{journal}{Phys. Rev. Lett.} \textbf{\bibinfo{volume}{78}},
  \bibinfo{pages}{1496} (\bibinfo{year}{1997}).

\bibitem[{\citenamefont{Salmon et~al.}(2003{\natexlab{b}})\citenamefont{Salmon,
  Manneville, Colin, and Pouligny}}]{Salmon:03_2}
\bibinfo{author}{\bibfnamefont{J.-B.} \bibnamefont{Salmon}},
  \bibinfo{author}{\bibfnamefont{S.}~\bibnamefont{Manneville}},
  \bibinfo{author}{\bibfnamefont{A.}~\bibnamefont{Colin}}, \bibnamefont{and}
  \bibinfo{author}{\bibfnamefont{B.}~\bibnamefont{Pouligny}},
  \bibinfo{journal}{Eur. Phys. J. AP} \textbf{\bibinfo{volume}{22}},
  \bibinfo{pages}{143} (\bibinfo{year}{2003}{\natexlab{b}}).

\bibitem[{\citenamefont{Salmon et~al.}(2003{\natexlab{c}})\citenamefont{Salmon,
  Manneville, and Colin}}]{Salmon:03_5}
\bibinfo{author}{\bibfnamefont{J.-B.} \bibnamefont{Salmon}},
  \bibinfo{author}{\bibfnamefont{S.}~\bibnamefont{Manneville}}, \bibnamefont{and}
  \bibinfo{author}{\bibfnamefont{A.}~\bibnamefont{Colin}},
  (\bibinfo{year}{2003}{\natexlab{c}}), \bibinfo{note}{submitted to Phys. Rev.
  E}.

\bibitem[{\citenamefont{Porte et~al.}(1991)\citenamefont{Porte, Appell,
  Bassereau, Marignan, Skouri, Billard, and Delsanti}}]{Porte:91}
\bibinfo{author}{\bibfnamefont{G.}~\bibnamefont{Porte}},
  \bibinfo{author}{\bibfnamefont{J.}~\bibnamefont{Appell}},
  \bibinfo{author}{\bibfnamefont{P.}~\bibnamefont{Bassereau}},
  \bibinfo{author}{\bibfnamefont{J.}~\bibnamefont{Marignan}},
  \bibinfo{author}{\bibfnamefont{M.}~\bibnamefont{Skouri}},
  \bibinfo{author}{\bibfnamefont{I.}~\bibnamefont{Billard}}, \bibnamefont{and}
  \bibinfo{author}{\bibfnamefont{M.}~\bibnamefont{Delsanti}},
  \bibinfo{journal}{Physica A} \textbf{\bibinfo{volume}{176}},
  \bibinfo{pages}{168} (\bibinfo{year}{1991}).

\bibitem[{\citenamefont{Roux et~al.}(1992)\citenamefont{Roux, Coulon, and
  Cates}}]{Roux:92}
\bibinfo{author}{\bibfnamefont{D.}~\bibnamefont{Roux}},
  \bibinfo{author}{\bibfnamefont{C.}~\bibnamefont{Coulon}}, \bibnamefont{and}
  \bibinfo{author}{\bibfnamefont{M.~E.} \bibnamefont{Cates}},
  \bibinfo{journal}{J. Phys. Chem.} \textbf{\bibinfo{volume}{96}},
  \bibinfo{pages}{4174} (\bibinfo{year}{1992}).

\bibitem[{\citenamefont{Herv\'e et~al.}(1993)\citenamefont{Herv\'e, Roux,
  Bellocq, Nallet, and Gulik-Krzywicki}}]{Herve:93}
\bibinfo{author}{\bibfnamefont{P.}~\bibnamefont{Herv\'e}},
  \bibinfo{author}{\bibfnamefont{D.}~\bibnamefont{Roux}},
  \bibinfo{author}{\bibfnamefont{A.-M.} \bibnamefont{Bellocq}},
  \bibinfo{author}{\bibfnamefont{F.}~\bibnamefont{Nallet}}, \bibnamefont{and}
  \bibinfo{author}{\bibfnamefont{T.}~\bibnamefont{Gulik-Krzywicki}},
  \bibinfo{journal}{J. Phys. II France} \textbf{\bibinfo{volume}{3}},
  \bibinfo{pages}{1255} (\bibinfo{year}{1993}).

\bibitem[{\citenamefont{Helfrich}(1978)}]{Helfrich:78}
\bibinfo{author}{\bibfnamefont{W.}~\bibnamefont{Helfrich}},
  \bibinfo{journal}{Z.~Naturforsch.} \textbf{\bibinfo{volume}{33a}},
  \bibinfo{pages}{305} (\bibinfo{year}{1978}).

\bibitem[{\citenamefont{Ackerson and Pusey}(1988)}]{Ackerson:88}
\bibinfo{author}{\bibfnamefont{B.~J.} \bibnamefont{Ackerson}} \bibnamefont{and}
  \bibinfo{author}{\bibfnamefont{P.~N.} \bibnamefont{Pusey}},
  \bibinfo{journal}{Phys. Rev. Lett.} \textbf{\bibinfo{volume}{61}},
  \bibinfo{pages}{1033} (\bibinfo{year}{1988}).

\bibitem[{\citenamefont{Wunenburger et~al.}(2001)\citenamefont{Wunenburger,
  Colin, Leng, Arn\'{e}odo, and Roux}}]{Wunenburger:01}
\bibinfo{author}{\bibfnamefont{A.-S.} \bibnamefont{Wunenburger}},
  \bibinfo{author}{\bibfnamefont{A.}~\bibnamefont{Colin}},
  \bibinfo{author}{\bibfnamefont{J.}~\bibnamefont{Leng}},
  \bibinfo{author}{\bibfnamefont{A.}~\bibnamefont{Arn\'{e}odo}},
  \bibnamefont{and} \bibinfo{author}{\bibfnamefont{D.}~\bibnamefont{Roux}},
  \bibinfo{journal}{Phys. Rev. Lett.} \textbf{\bibinfo{volume}{86}},
  \bibinfo{pages}{1374} (\bibinfo{year}{2001}).

\bibitem[{\citenamefont{Salmon et~al.}(2002)\citenamefont{Salmon, Colin, and
  Roux}}]{Salmon:02}
\bibinfo{author}{\bibfnamefont{J.-B.} \bibnamefont{Salmon}},
  \bibinfo{author}{\bibfnamefont{A.}~\bibnamefont{Colin}}, \bibnamefont{and}
  \bibinfo{author}{\bibfnamefont{D.}~\bibnamefont{Roux}},
  \bibinfo{journal}{Phys. Rev. E} \textbf{\bibinfo{volume}{66}},
  \bibinfo{pages}{031505} (\bibinfo{year}{2002}).

\bibitem[{\citenamefont{Grosso et~al.}(2001)\citenamefont{Grosso, Keunings,
  Crescitelli, and Maffettone}}]{Grosso:01}
\bibinfo{author}{\bibfnamefont{M.}~\bibnamefont{Grosso}},
  \bibinfo{author}{\bibfnamefont{R.}~\bibnamefont{Keunings}},
  \bibinfo{author}{\bibfnamefont{S.}~\bibnamefont{Crescitelli}},
  \bibnamefont{and} \bibinfo{author}{\bibfnamefont{P.~L.}
  \bibnamefont{Maffettone}}, \bibinfo{journal}{Phys. Rev. Lett.}
  \textbf{\bibinfo{volume}{86}}, \bibinfo{pages}{3184} (\bibinfo{year}{2001}).

\bibitem[{\citenamefont{Cates et~al.}(2002)\citenamefont{Cates, Head, and
  Ajdari}}]{Cates:02}
\bibinfo{author}{\bibfnamefont{M.~E.} \bibnamefont{Cates}},
  \bibinfo{author}{\bibfnamefont{D.~A.} \bibnamefont{Head}}, \bibnamefont{and}
  \bibinfo{author}{\bibfnamefont{A.}~\bibnamefont{Ajdari}},
  \bibinfo{journal}{Phys. Rev. E} \textbf{\bibinfo{volume}{66}},
  \bibinfo{pages}{025202} (\bibinfo{year}{2002}).

\bibitem[{\citenamefont{Berne and Pecora}(1995)}]{Berne:95}
\bibinfo{author}{\bibfnamefont{B.~J.} \bibnamefont{Berne}} \bibnamefont{and}
  \bibinfo{author}{\bibfnamefont{R.}~\bibnamefont{Pecora}},
  \emph{\bibinfo{title}{Dynamic light scattering}} (\bibinfo{publisher}{Wiley,
  New York}, \bibinfo{year}{1995}).

\bibitem[{\citenamefont{Ackerson and Clark}(1981)}]{Ackerson:81}
\bibinfo{author}{\bibfnamefont{B.~J.} \bibnamefont{Ackerson}} \bibnamefont{and}
  \bibinfo{author}{\bibfnamefont{N.~A.} \bibnamefont{Clark}},
  \bibinfo{journal}{J. Physique} \textbf{\bibinfo{volume}{42}},
  \bibinfo{pages}{929} (\bibinfo{year}{1981}).

\bibitem[{\citenamefont{Fuller et~al.}(1980)\citenamefont{Fuller, Rallison,
  Schmidt, and Leal}}]{Fuller:80}
\bibinfo{author}{\bibfnamefont{G.~G.} \bibnamefont{Fuller}},
  \bibinfo{author}{\bibfnamefont{J.~M.} \bibnamefont{Rallison}},
  \bibinfo{author}{\bibfnamefont{R.~L.} \bibnamefont{Schmidt}},
  \bibnamefont{and} \bibinfo{author}{\bibfnamefont{L.~G.} \bibnamefont{Leal}},
  \bibinfo{journal}{J. Fluid Mech.} \textbf{\bibinfo{volume}{100}},
  \bibinfo{pages}{555} (\bibinfo{year}{1980}).

\bibitem[{\citenamefont{Maloy et~al.}(1992)\citenamefont{Maloy, Goldburg, and
  Pak}}]{Maloy:92}
\bibinfo{author}{\bibfnamefont{K.~J.} \bibnamefont{M\aa l\o y}},
  \bibinfo{author}{\bibfnamefont{W.}~\bibnamefont{Goldburg}}, \bibnamefont{and}
  \bibinfo{author}{\bibfnamefont{H.~K.} \bibnamefont{Pak}},
  \bibinfo{journal}{Phys. Rev. A} \textbf{\bibinfo{volume}{46}},
  \bibinfo{pages}{3288} (\bibinfo{year}{1992}).

\bibitem[{\citenamefont{Gollub and Freilich}(1974)}]{Gollub:74}
\bibinfo{author}{\bibfnamefont{J.~P.} \bibnamefont{Gollub}} \bibnamefont{and}
  \bibinfo{author}{\bibfnamefont{M.~H.} \bibnamefont{Freilich}},
  \bibinfo{journal}{Phys. Rev. Lett.} \textbf{\bibinfo{volume}{33}},
  \bibinfo{pages}{1465} (\bibinfo{year}{1974}).

\bibitem[{\citenamefont{Radulescu and Olmsted}(2000)}]{Radulescu:00}
\bibinfo{author}{\bibfnamefont{O.}~\bibnamefont{Radulescu}} \bibnamefont{and}
  \bibinfo{author}{\bibfnamefont{P.~D.} \bibnamefont{Olmsted}},
  \bibinfo{journal}{J.~Non-Newtonian Fluid Mech.}
  \textbf{\bibinfo{volume}{91}}, \bibinfo{pages}{143} (\bibinfo{year}{2000}).

\bibitem[{\citenamefont{Barnes}(1995)}]{Barnes:95}
\bibinfo{author}{\bibfnamefont{H.~A.} \bibnamefont{Barnes}},
  \bibinfo{journal}{J. Non-Newtonian Fluid Mech.}
  \textbf{\bibinfo{volume}{56}}, \bibinfo{pages}{221} (\bibinfo{year}{1995}).

\bibitem[{\citenamefont{Salmon et~al.}(2003{\natexlab{d}})\citenamefont{Salmon,
  B\'ecu, Manneville, and Colin}}]{Salmon:03_1}
\bibinfo{author}{\bibfnamefont{J.-B.} \bibnamefont{Salmon}},
  \bibinfo{author}{\bibfnamefont{L.}~\bibnamefont{B\'ecu}},
  \bibinfo{author}{\bibfnamefont{S.}~\bibnamefont{Manneville}},
  \bibnamefont{and} \bibinfo{author}{\bibfnamefont{A.}~\bibnamefont{Colin}},
  \bibinfo{journal}{Eur. Phys. J. E} \textbf{\bibinfo{volume}{10}},
  \bibinfo{pages}{209} (\bibinfo{year}{2003}{\natexlab{d}}).

\bibitem[{\citenamefont{Welch et~al.}(2002)\citenamefont{Welch, Stetzer, Hu,
  Sirota, and Idziak}}]{Welch:02}
\bibinfo{author}{\bibfnamefont{S.~E.} \bibnamefont{Welch}},
  \bibinfo{author}{\bibfnamefont{M.~R.} \bibnamefont{Stetzer}},
  \bibinfo{author}{\bibfnamefont{G.}~\bibnamefont{Hu}},
  \bibinfo{author}{\bibfnamefont{E.~B.} \bibnamefont{Sirota}},
  \bibnamefont{and} \bibinfo{author}{\bibfnamefont{S.~H.~J.}
  \bibnamefont{Idziak}}, \bibinfo{journal}{Phys. Rev. E}
  \textbf{\bibinfo{volume}{65}}, \bibinfo{pages}{061511}
  (\bibinfo{year}{2002}).

\end{thebibliography}

\end{document}